\documentclass[final,5p,times,twocolumn]{elsarticle}

\usepackage{amssymb}
\usepackage{amsmath}
\usepackage{lineno}
\usepackage{graphicx} % Required for inserting images

\usepackage[colorlinks=true,breaklinks=true]{hyperref}
\usepackage{changepage} % Para el entorno adjustwidth
\usepackage{tabularx} % Para tablas ajustables
\usepackage{makecell} %para partir lineas en celdas de tablas
\usepackage{pifont} % More styles for bullets
\usepackage{caption}
\usepackage{subcaption}

\usepackage[dvipsnames]{xcolor}

\usepackage[colorlinks=true,breaklinks=true]{hyperref}
\usepackage[normalem]{ulem}
\usepackage[utf8]{inputenc}
\hypersetup{allcolors=[rgb]{0.0 0.0 0.6},linkcolor=[rgb]{0.75 0.05 0.05}}
\usepackage{amsmath,amssymb}
\usepackage{epsfig}  
\usepackage{graphicx}   
\usepackage{slashed}       
\usepackage{tikz}
\usepackage{color}
\usepackage{tikz-feynman}
\usepackage{url}
\usepackage{color}
\usepackage{multirow}
\usepackage{comment}
\usepackage{amssymb}
\usepackage{bm}
\usepackage{footmisc}

\newcommand{\beq}{\begin{equation}}
\newcommand{\eeq}{\end{equation}}

\definecolor{forestgreen}{rgb}{0.13, 0.545, 0.13}

\newcommand{\cfs}[1]{{\color{magenta}[CFS: #1]}}

\allowdisplaybreaks

\bibpunct{[}{]}{,}{n}{}{,}

\journal{Physics of the Dark Universe}

\begin{document}

\begin{frontmatter}

\title{Deeper analysis of \emph{Fermi}-LAT unassociated 4FGL J2112.5-3043 for possible identification}

\author[inst1,inst2]{Federica Giacchino\corref{cor1}}
\ead{federica.giacchino@roma2.infn.it}
\cortext[cor1]{Corresponding author.}

\affiliation[inst1]{organization={Department of Fundamental Physics, University of Salamanca},
addressline={Plaza de la Merced s/n},
city={Salamanca},
postcode={E-37008},
country={Spain}}

%\affiliation[inst1]{Department of Fundamental Physics, University of Salamanca, Plaza de la Merced s/n, E-37008 Salamanca, Spain.}
\affiliation[inst2]{organization={Istituto Nazionale di Fisica Nucleare - Sezione Roma Tor Vergata},
addressline={via della Ricerca Scientifica},
city={Roma},
postcode={00133},
country={Italy}}

%\affiliation[inst2]{Istituto Nazionale di Fisica Nucleare - Sezione Roma Tor Vergata, via della Ricerca Scientifica, 00133 Roma, Italy.}

\author[inst3,inst4]{Cristina Fernández-Suárez}
\ead{cristina.fernandezs01@estudiante.uam.es}
\affiliation[inst3]{organization={Instituto de Fisica Teorica UAM-CSIC, Universidad Autonoma de Madrid}, %Department and Organization
            addressline={C Nicolas Cabrera, 13-15}, 
            city={Madrid},
            postcode={28049}, 
         %   state={Madrid},
            country={Spain}}
\affiliation[inst4]{organization={Departamento de Fisica Teorica, M-15, Universidad Autonoma de Madrid},%Department and Organization
      %      addressline={}, 
            city={Madrid},
            postcode={E-28049}, 
       %     state={},
            country={Spain}}

\author[inst3,inst4]{Miguel \'A. S\'anchez-Conde}
\ead{miguel.sanchezconde@uam.es}

\author[inst1]{M.\'Angeles P\'erez-Garc\'ia}
\ead{mperezga@usal.es}

\author[inst2,inst5]{Stefano Ciprini}
\ead{stefano.ciprini@roma2.infn.it}

\affiliation[inst5]{organization={ASI Space Science Data Center},
addressline={via del Politecnico},
city={Roma},
postcode={00133},
country={Italy}}

%\affiliation[inst5]{ASI Space Science Data Center, via del Politecnico, 00133, Roma, Italy}

\author[inst2,inst5]{Dario Gasparrini}
\ead{dario.gasparrini@roma2.infn.it}


%\linenumbers
%% Abstract

\begin{abstract}

In the 4FGL-DR4 point-source catalog of the Large Area Telescope (LAT) onboard NASA’s \emph{Fermi} Gamma-ray Observatory (\emph{Fermi}-LAT), around a third of the sources are still unidentified (unIDs). In this work, we perform a detailed study of one of them, namely 4FGL J2112.5-3043. Only gamma-ray emission has been detected from this unidentified source, with no counterpart observed at any other wavelength as of today. Together with its high detection significance, this makes 4FGL J2112.5-3043 a particularly compelling target for further investigation. %\apg{Incidentally, we comment on a 162 TeV neutrino reported by IceCube detector close to the source direction.} 
The results of our spectral and spatial analyses show that the source photon spectrum is better described with a subexponential cutoff power-law spectral model, with no significant flux variability over time, and a morphology consistent with being a point-like source. We investigate and discuss the characterized emission within the context of both conventional and exotic astrophysics, namely a pulsar origin or potential dark matter (DM) annihilations in a nearby Galactic subhalo. Although our results are inconclusive and neither confirm a DM origin nor firmly establish an astrophysical nature, we find a spectral preference for the $b\bar{b}$ and $c\bar{c}$ DM annihilation channels over a pulsar origin, thus making this unID a particularly intriguing candidate for next multiwavelength observations.

\end{abstract}

%% Keywords

\end{frontmatter}

\section{Introduction}

The Large Area Telescope (LAT, \cite{Gehrels:1999ri, Fermi-LAT:2009ihh}) onboard the \emph{Fermi} Gamma-ray Space Telescope is a pair-conversion telescope, put in orbit in 2008, designed to detect photons in the energy range from 20 MeV to $>$ 1 TeV. In the latest \emph{Fermi}-LAT Fourth Source Catalog Data Release 4 (4FGL-DR4~\citep{Ballet:2023qzs}), $\sim 2500$ objects still remain as unidentified sources (unIDs), i.e. these objects have no clear single association or multiwavelength counterpart, to either a known object identified at other wavelengths, or to a known source type emitting only in gamma rays (such as certain pulsars).

This opens the possibility that some of these unIDs may not have a purely astrophysical origin, but rather be related to new physics, in particular, to dark matter (DM) (e.g.,~\cite{Ackermann_2012, Coronado-Blazquez:2019puc, Coronado-Blazquez:2019pny, Coronado-Blazquez:2021amj}). There is compelling evidence that 85$\%$ of the matter in the Universe is in the form of DM, although its nature is still unknown~\cite{Bertone_2005, Bertone_2018, Balazs:2024uyj}. In this context, indirect searches for DM  looking for the annihilation or decay products of DM particles in the form of photons at gamma-ray energies, antimatter and/or neutrinos may be key to helping us find it. Several works involving gamma-ray telescopes, such as the \emph{Fermi}-LAT, have explored this scenario trying to look for a DM hint coming from different astrophysical targets, including the Galactic center (e.g.,~\cite{Ajello_2016, Ackermann_2017, Di_Mauro_2021}), galaxy clusters (e.g.,~\cite{Ackermann_2015, Di_Mauro_2023}), dwarf spheroidal galaxies (dSphs, e.g.,~\cite{Fermi-LAT:2015att, armand2021combineddarkmattersearches, mcdaniel2023legacyanalysisdarkmatter, Fermi-LAT:2025gei}), stellar streams (\cite{Fernandez-Suarez_2025}), or dark satellites (e.g.,~\cite{Ackermann_2012, Coronado-Blazquez:2019puc, Coronado-Blazquez:2019pny, Coronado-Blazquez:2021amj}). %~\cite{Madau:2008fr}). 
The latter, DM subhalos devoid of baryons, would have no visible counterparts and may appear as unIDs in gamma-rays. However, no unID source has yet been conclusively confirmed to originate from DM. Thus, a detailed case-by-case study of unIDs is essential to clarify their origin and assess the possibility that some of them may actually be related to DM.

As a starting point in this work, we have ranked the 4FGL-DR4 unIDs by detection significance, excluding those within the Galactic plane ($|b| \leq 3$). This exclusion is motivated by the high concentration of astrophysical sources -- such as pulsars and supernova remnants -- near the Galactic plane, making source identification in this region particularly challenging. Furthermore, since dark satellites are not expected to emit at wavelengths other than gamma rays, we selected only those sources with no detected multiwavelength counterparts within twice the error circle ($2ec$) of the \textit{Fermi}-LAT in the ASI Space Data Center (SSDC). We then searched for potential counterparts or multifrequency emitters in the vicinity of the unIDs using the NED\footnote{https://ned.ipac.caltech.edu/} and SIMBAD\footnote{ http://simbad.u-strasbg.fr/simbad/} databases, followed by a systematic review of the literature. After this filtering process, we identified  4FGL~J2112.5$-$3043 as the highest-significance unID that satisfies all selection criteria. Consequently, we carried out a detailed analysis of this object to assess whether it is compatible with a DM origin or, alternatively, more likely associated with an astrophysical source. Interestingly, in ref.~\citep{Mirabal:2021ayb}, authors provide a short list of \textit{Fermi} unIDs most compatible with being dark satellites, including our unID of interest, 4FGL J2112.5-3043. Among that list, they highlight 4FGL J2112.5-3043 as a promising dark satellite, even potentially responsible for the perturbation observed in the GD-1 stellar stream, and emphasize the need for a more detailed analysis.

This work is organized as follows: we show the multifrequency research performed for 4FGL J2112.5-3043 in Sec.~\ref{sec:4FGLJ2112.5-3043}. Then, we present the \emph{Fermi}-LAT data analysis along with the main results in Sec.~\ref{sec:data analysis}. We next discuss the possible interpretations of our findings in Sec.~\ref{sec:intepretation}, considering both the potential hints of new physics and the more conventional explanation in terms of a pulsar origin. Finally, we present our conclusions in Sec.~\ref{sec:conclusions}.

\section{Multifrequency search of 4FGL J2112.5-3043}
\label{sec:4FGLJ2112.5-3043}

4FGL 2112.5-3043 is located at $\mathrm{R.A.}=318.14$ deg,  $\mathrm{Dec.}=-30.729$ deg. According to the 4FGL-DR4 catalog~~\citep{Ballet:2023qzs}, it is a bright source with an integrated flux of $\Phi(1<E<100 \text{ GeV})=(3.49 \pm 0.10)\times 10^{-9}$ ph/cm$^2$/sec and a Test Statistic (TS, see Eq. \ref{eq:ts} in Sec. \ref{sec:data analysis}) of $\mathrm{TS}\sim4170$, which corresponds to a detection significance of $\sim 65$ (in sigma units). Moreover, the variability index computed in 1-year is $\sim$ 14.93, thus showing no evidence of variability\footnote{
Using the same method applied in~\cite{Fermi-LAT:2022byn}, the variability threshold is defined as the $\chi^{2}$ critical value for $N-1$ degrees of freedom, where $N$ is the number of time intervals. In the 4FGL-DR4 catalog, a source is considered to be variable when its variability index exceeds this value; for example, for $N=13\,(18)$ intervals, the 99\% confidence threshold is $\mathrm{TS}_{\mathrm{var}} = 27.69 \,(33.4)$.}.

The first step in assessing whether our unID could be compatible with a dark satellite is to investigate whether it emits at wavelengths other than gamma rays, as these objects are devoid of baryons and thus are not expected to show multiwavelength counterparts. To this end, we carried out a detailed review on the literature and catalogs of major experiments to search for any possible associations at other wavelengths. The error circle ($ec$) of the \emph{Fermi}-LAT, defined as %given as 
the semi-minor axis of the error ellipse at $95\%$ confidence level (C.L.), is $ec\sim 1.07$ arcmins. Within a radius of $2ec$, no other relevant detections are found in the SSDC database, nor in the SED Builder, NED, or SIMBAD catalogs. Next, we present in detail the main results of our multifrequency research.
%The 4FGL-DR4 integrated flux is $\Phi(1<E<100 \text{ GeV})=(3.49 \pm 0.10)\times 10^{-9}$ ph/cm$^2$/sec.

%The first step in assessing whether our unID could be compatible with a DM subhalo is to investigate whether it emits at wavelengths other than gamma rays, as dark objects are not expected to show multiwavelength counterparts. To this end, we carried out a detailed review of the literature and catalogs of major experiments to search for any possible associations at other wavelengths.

\subsection{X-ray observations}
Millisecond pulsars (MSPs) can produce an X-ray emission from the pulsar magnetosphere and/or from the shock \cite{2009ApJ...699.1171A}. 
Our source of interest, 4FGL 2112.5-3043, is X-ray covered by the XMM-Newton~\cite{Struder:2001bh, Turner:2000jy}, Swift~\cite{SWIFT:2005ngz} and Chandra~\cite{Elvis:2009rr} experiments. In ref.~\citep{Salvetti:2017pwl}, three very faint X-ray observations (near the detection threshold) have been detected by XMM-Newton within the 3FGL error circle (considering the previous LAT data release, 4FGL-DR3 \cite{Fermi-LAT:2022byn}, one of them is completely outside this region). These sources are described with a power-law spectrum model and show no optical counterpart in Catalina~\cite{Drake:2008vy}, nor any significant periodic modulation or short-term variability. 
We have found only one faint source, 4XMM J211232.1-304403, with a flux of $\Phi=(1.54\pm 0.43)\times 10^{-14}$ mW/m$^2$ in the $(0.2-12)$ keV energy band. 
This detection is localized at $\mathrm{R.A.}=318.131$ deg, $\mathrm{Dec.}=-30.734$ deg, with an error radius of $ec = 1.074$ arcmins and lying approximately $1.1$ arcmin from our unID. 
%with the γ-ray unID due to the lack of variability or pulsations and the absence of multiwavelength counterparts.
However, as reported in ref.~\citep{Salvetti:2017pwl}, it cannot be associated with our unID due to the lack of variability or pulsations. %and the absence of multiwavelength counterpart}. 
Therefore, under a pulsar or MSP interpretation of our unID, this X-ray source would not constitute a viable counterpart.
Finally, in ref.~\citep{Mirabal:2021ayb} a new X-ray analysis is performed in the error region of our unID, only providing an upper limit ($< 2.5 \times 10^{-14}$ erg cm$^{-2}$ s$^{-1}$), since no Swift detection has been found.  

 \subsection{Infrared observations}

Ref.~\citep{Coronado-Blazquez:2019puc} analysed several unassociated sources in the 2FHL \cite{Fermi-LAT:2015uah}, 3FHL \cite{Fermi-LAT:2017sxy}, and 3FGL \cite{Fermi-LAT:2015bhf} \emph{Fermi} catalogs. Our unID candidate (referred to there as 3FGL 2112.5-3044) was excluded because ref.~\citep{Pena-Herazo:2017evn} associated it with the infrared source WISE J211217.41-304655.3, for which they performed an optical spectroscopic analysis on a sample of $\sim 60$ blazar-like targets. In their work, the selection relies on an infrared gamma-ray connection traced through WISE colors~\cite{Massaro:2012pi}.
However, the angular separation between this WISE source and the 3FGL 2112.5-3044 position is $\sim 4.6,\text{arcmin}$, which is greater than $2ec$, making this association unlikely.

\subsection{Other observations}

In the optical band, $\sim30$ objects have been identified within the error circle of our unID using the Gaia DR3 catalog~\cite{GaiaDR3}, with no evidence of variability. Ref.~\cite{Braglia:2020jri} conducted an optical search for binary companions, finding no significant detections. Additional observations with ULTRACAM\footnote{\url{https://www.eso.org/public/images/2017_11_16_La_Silla_NTT_ULTRACAM_upr_IMG_2110-CC/}} are planned to follow up on this search.

In terms of radio frequencies, no counterpart has been observed, despite multiple radio observatories targeting the region of interest.

Other intriguing multimessenger signals have also been reported. In particular, IceCube has detected the alert track event IC181212A, with a reconstructed neutrino energy of $E \sim 162$ TeV. Our unID falls within the 90\% C.L. region of the neutrino~\citep{IceCube:2023agq}.
%(see Fig.~\ref{fig:neutrinoposition}).
Typically, such energetic neutrinos are believed to be produced in hadronic reactions from accelerated PeV-protons $p+p \rightarrow \pi^{ \pm} \rightarrow \mu^{ \pm}+\nu_\mu \rightarrow e^{ \pm}+\nu_e+\nu_\mu$. 
Assuming a distance of $d = 10$ kpc for a hypothetical supernova remnant hosting a highly magnetized pulsar, we can roughly estimate the neutrino luminosity as $L_\nu=\mathcal{F}_\nu \times 4 \pi d^2 \approx 3.1 \times 10^{41} \mathrm{erg} / \mathrm{s}$. 
%[Considering proton spectral index emission $\frac{d N_p}{d E_p} \sim E_p^{-s}$ with spectral index $s\sim 2$ we obtain a fraction of them suitable to produce such event $f_{E, p>1 \mathrm{PeV}} \approx 0.14$, so that their total energy is $E_{p>1 \mathrm{PeV}} \approx 1.43 \times 10^{49} \mathrm{erg}$ so that the efficiency and and $\varepsilon_{p p}=0.1$  $k_\nu=0.05$ ) so that the total energy into neutrinos is $E_{\nu, \text { total }} \approx 1 \times %10^{47} \mathrm{erg}$. This allowing a normalized $N_{\nu, \mathrm{dec}} \approx 6.8 \times 10^{32}$ and a detected number of neutrinos at IceCube $N_{\mathrm{det}}=N_{\nu, \mathrm{dec}} \times \frac{A_{\mathrm{eff}}}{4 \pi d^2}$ assuming effective area for IceCube at that energy region
%$A_{\text {eff }}=10^{10} \mathrm{~cm}^2$. Just for giving an idea, at $10\, \mathrm{kpc}$ one should expect $N_{\mathrm{det}} \approx 5.7 \times 10^{-4}$.]
For a proton spectrum $\frac{d N_p}{d E_p} \sim E_p^{-s}$ with spectral index $s\sim 2$, the fraction of protons above 1 PeV suitable to produce such event is $
f_{E, p>1 \mathrm{PeV}} \approx 0.14$, leading to an expected number of detected neutrinos in IceCube of $N_{\mathrm{det}} \approx 5.7 \times 10^{-4}$.
Hence, such a detection is a very rare — though not impossible — event. We also note that if the source were extragalactic, the expected $N_{\mathrm{det}}$ would be even smaller. 

As a comparison, if we consider whether this neutrino event could potentially be associated with new physics arising from DM self-annihilation,  we find that such an interpretation is %energetically forbidden. 
disfavoured in the WIMP vanilla scenario \citep{Balazs_2026}.
However, more elaborated DM models, in which the dark candidate acts as a cosmic-ray target or produces a boosted component, could in principle accommodate such high-energy neutrinos~\citep{Bell:2021zkr,Agashe:2014yua}. Exploring these possibilities lies beyond the scope of this paper and is left for future work.

%\begin{figure}[h]
%\begin{center}{\includegraphics[height=0.85\linewidth]{plot neutrino vs Fermi.png}}
       % \caption{Neutrino event (red point) at $90\%$ C.L. (red dashed line) and unID 4FGL 2112.5-3043 position (blue point) with $0.1^{\circ}$ as error region from the \textit{Fermi}-LAT point spread function (PSF) at $10$ GeV (blue dashed line). \cfs{I suggest that the X-axis and Y-axis labels be the same size, and that the legend size be larger} \cfs{Miguel suggests removing this plot from the paper.}}
%\label{fig:neutrinoposition}
%\end{center}
%\end{figure}
%\fg{This neutrino tells something? Some WIMP analysis?}

\section{\textit{Fermi}-LAT Data Analysis}
\label{sec:data analysis}

In this section, we perform a dedicated analysis of 4FGL J2112.5-3043. We first describe our analysis methodology and then present our main results, including the spectral, variability, and spatial extension analyses of this source.

\subsection{Technical setup and analysis methodology}

\begin{table}[t]
\begin{center}
\begin{tabular*}{\linewidth}{l@{\extracolsep{\fill}}c}
\hline
\hline
Parameter name & Value \\
\hline
\hline
 \multirow{2}{10em}{Time Domain ISO8601 (UTC)} & from 2008-08-04 15:43:36 \\
  & to 2025-09-01 07:47:48 \\
 \hline
 Time Domain (MET) & from 239557417 to 778405673 \\ 
 \hline
 Energy Range & $[0.1-1000]$ GeV \\
 \hline
 IRF & $\text{P8R3}\_\text{SOURCE}\_\text{V3}$ \\
 \hline
 Event Type &  FRONT + BACK \\
 \hline
 Point-Source Catalog &  4FGL-DR4 \\ 
 \hline
 ROI size & $15^{\circ}\times 15^{\circ}$ \\
 \hline
 Angular bin size & $0.1^{\circ}$  \\ 
 \hline
 Bins per energy decade & 8 \\
 \hline
 Galactic diffuse model &  $\text{gll}\_\text{iem}\_\text{v07}.\text{fits}$ \\ 
 \hline
 Isotropic diffuse model &  $\text{iso}\_\text{P8R3}\_\text{SOURCE}\_\text{V3}\_\text{v1}.\text{txt}$ \\ 
\hline
\end{tabular*}
\caption{Description of the \emph{Fermipy} analysis technical setup.}
\label{tab:summary}
\end{center}
\end{table}

The LAT data analysis of our unID 4FGL J2112.5-3043 was performed with \emph{Fermipy}\footnote{\url{https://fermipy.readthedocs.io/en/latest/}} (version 1.2.0)~\citep{Wood:2017yyb}, which uses the underlying Fermitools\footnote{\url{https://fermi.gsfc.nasa.gov/ssc/data/analysis/software/}} software packages (version 2.2.0). We have used $17$ years of data Mission Elapsed Time (MET): from
239557417 to 778405673 (ISO8601 UTC: 2008-08-04 15:43:36, 2025-09-01 07:47:48) in the energy range $[0.1-1000]$ GeV, using Pass 8 events~\cite{Fermi-LAT:2013jgq, Bruel:2018lac} and all the available photons (\texttt{FRONT + BACK}) of the SOURCE class, excluding those arriving with zenith angles greater than $90^{\circ}$ to get rid of the contamination from the Earth's atmosphere. We also apply the \texttt{$(\text{DATA}\_\text{QUAL}>0)\&\&(\text{LAT}\_\text{CONFIG}==1)$} filter to ensure data quality. The data have been taken within a region of interest (ROI) of $15^{\circ}\times 15^{\circ}$ around the unID, with a pixel size of $0.1^{\circ}$ and $8$ evenly spaced logarithmic energy bins. We have built the sources model database, based on the fourth incremental version of the fourth \textit{Fermi}-LAT point-source catalog (4FGL-DR4~\citep{Ballet:2023qzs}). We used the \texttt{$\text{P8R3}\_\text{SOURCE}\_\text{V3}$} instrumental response functions (IRFs), along with the Galactic diffuse emission model \texttt{$\text{gll}\_\text{iem}\_\text{v07}.\text{fits}$} and the isotropic component \texttt{$\text{iso}\_\text{P8R3}\_\text{SOURCE}\_\text{V3}\_\text{v1}.\text{txt}$}\footnote{\url{https://fermi.gsfc.nasa.gov/ssc/data/access/lat/BackgroundModels.html}}. We summarize the setup parameters in Table~\ref{tab:summary}. 

We set up the analysis by running the \texttt{GTAnalysis} and \texttt{GTAnalysis.setup} modules. Next, we perform an automatic optimization of the ROI with \texttt{GTAnalysis.optimize} and free the normalization and spectral shape of all the sources within $3^{\circ}$ of the ROI center with \texttt{GTAnalysis.free}, as well as the normalization and spectral index of the Galactic diffuse component, and the normalization of the isotropic diffuse template. After leaving free all these parameters, we re-optimize the model with the \texttt{GTAnalysis.fit} module.

The detection significance of the source is defined by the Test Statistic (TS) as:
\begin{equation}
    \mathrm{TS} = 2 \, \log \left(\frac{\mathcal{L} (H_1)}{\mathcal{L} (H_0)} \right),
    \label{eq:ts}
\end{equation}
where $\mathcal{L} (H_0)$ and $\mathcal{L} (H_1)$ represent the likelihoods under the null (no  source) hypothesis and the alternative (existing source) hypothesis, respectively. It is common to set the detection threshold at $\mathrm{TS}=25$, which corresponds to a detection significance of $\sim 5 \sigma$ (being $\sigma$ the significance expressed in standard deviations)\footnote{In an ideal scenario, the TS values obtained from a gamma-ray analysis are expected to follow a $\chi^2$ distribution. Based on the Chernoff's theorem \cite{chernoff}, the TS distribution can be well-approximated by a $\chi^2$ distribution with \textit{n} degrees of freedom divided by two. This approximation enables a straightforward relation between the TS and the detection significance as $\sqrt{\mathrm{TS}} \sim \sigma$.}, since this is the usual threshold above which sources are included in the 4FGL-DR4 catalog. The detection TS obtained in our analysis of 4FGL J2112.5-3043 is $\mathrm{TS}=4756.46 \, (\sim 69 \,\sigma)$, being slightly higher than the value provided in the 4FGL-DR4 catalog ($\mathrm{TS}=4170$) due to the increase in the observation period.

\subsection{Analysis results}

\begin{figure}
    \centering
    \includegraphics[width=1.0\linewidth]{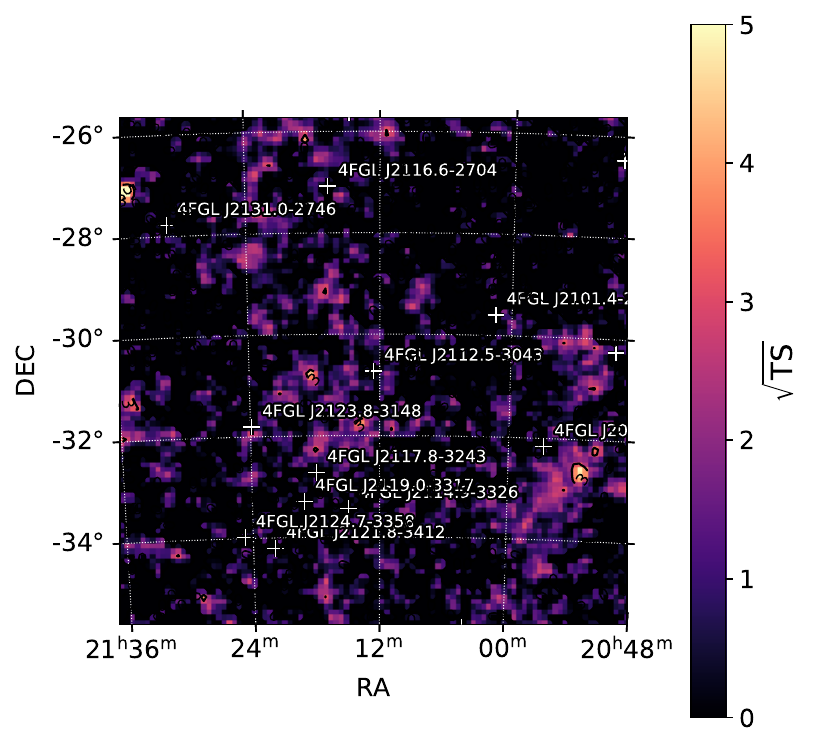}
    \caption{Significance map (i.e., $\sqrt{TS}$, see Eq. \ref{eq:ts}) for the $15^{\circ} \times 15^{\circ}$ ROI around 4FGL 2112.5-3043. White crosses indicate the positions of point sources in the model, in the energy range between $100$ MeV and $1$ TeV.}
    \label{TSMAP}
\end{figure}
\begin{figure}[h]
\begin{center}
        {\includegraphics[height=0.73\linewidth]{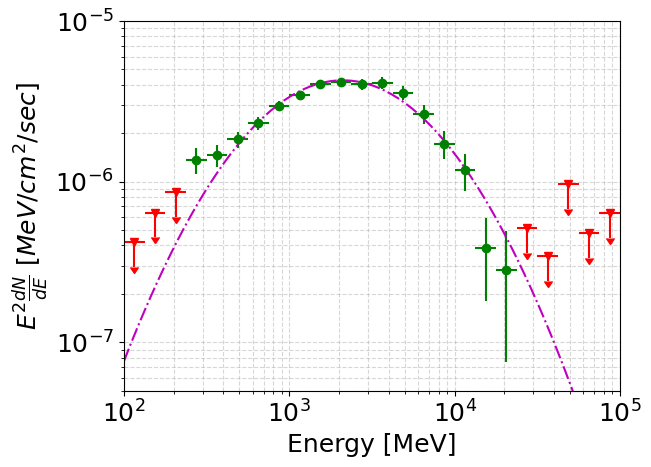}}
        \caption{Gamma-ray energy spectrum of our unID 4FGL 2112.5-3043, where the green points correspond to a signal detection ($\mathrm{TS} \geq 4$), while the red ones show the 95\% C.L. upper limits to the flux, reported when $\mathrm{TS} < 4$. The best-fit spectral model described by a log-parabola is also shown (dot-dashed purple line).}
\label{fig:spectrum}
\end{center}
\end{figure}

After the point-source analysis of our unID performed with the aforementioned technical setup, we obtain the TS map of the $15^{\circ} \times 15^{\circ}$ ROI around 4FGL J2112.5-3043 by means of the \texttt{GTAnalysis.tsmap} routine, as displayed in Fig.~\ref{TSMAP}. This map shows the $\sqrt{TS}$ of the residual emission after modeling all sources in the ROI, and therefore does not correspond to the TS of the target source itself. The gamma-ray spectral energy distribution (SED) of the source is generated with the \texttt{GTAnalysis.sed} module using $8$ bins per decade, as shown in Fig.~\ref{fig:spectrum}. 
The best-fit (purple dot-dashed line in Fig.~\ref{fig:spectrum}) to the data provided by the analysis is a log-parabola (LP) spectral model:
\begin{equation}
    \frac{dN}{dE}=\phi_0\Big(\frac{E}{E_0}\Big)^{-\alpha-\beta \,Log(\frac{E}{E_0})},
    \label{equation:LP}
\end{equation}
where the values of the spectral parameters are generated by the \texttt{GTAnalysis.sed} module, namely: the normalization prefactor $\phi_0=(2.06\,\pm\, 0.06)\times 10^{-12}$ MeV$^{-1}$cm$^{-2}$ sec$^{-1}$, the slope index $\alpha=1.64\pm0.04$, the spectral curvature index $\beta=0.43\pm0.03$, and, finally, the scale parameter $E_0=1386$ MeV. This parametric function is often used to fit curved spectra, such as those from pulsars, although it is also a good description of the spectrum expected by certain DM annihilation channels \citep{Coronado-Blazquez:2019pny}. %\footnote{As an additional check, we run the curvature test with the \texttt{GTAnalysis.curvature} module for the source of interest, in order to test for spectral curvature (deviation from a power law (PL) energy spectrum). We have obtained TS$_{\text{curv}} = 457.646$ for a log-parabola and TS$_{\text{curv}} =466.121$ for a power law with super-exponential cutoff (PLSC) with a super-exponential index of $0.667$, indicating a slight preference towards a PLSC spectrum.}. 
The upper limits are set for $\mathrm{TS} < 4$, since the source position is known and fixed, and a $2\sigma$ confidence level is considered sufficient.

%\fg{Say something about some green points do not fit well with LP? Other component at low energies?}

\begin{figure}[h]
\begin{center}        {\includegraphics[height=0.73\linewidth]{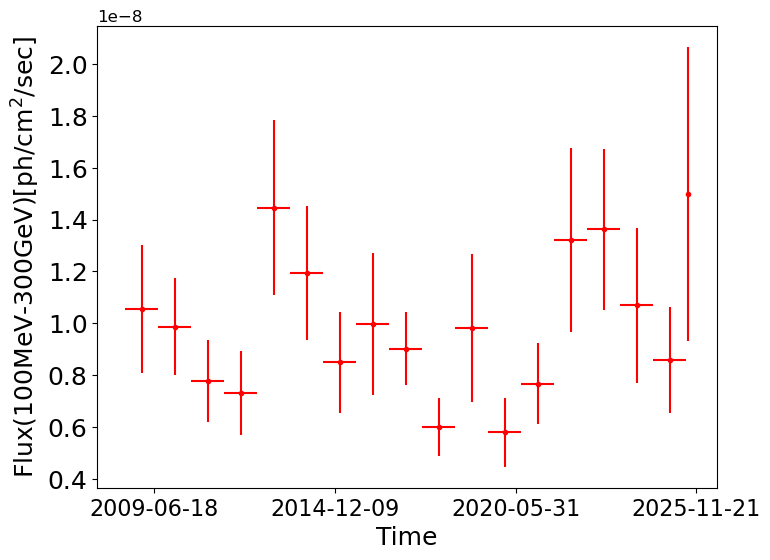}}
        \caption{Light Curve of 4FGL 2112.5-3043 in $17$ years of LAT data, with a time binning of one year.}
\label{fig:lc}
\end{center}
\end{figure}

Next, we study the variability of the source, since DM annihilation in subhalos is expected to be steady, i.e., their flux should not change over time. We extracted its light curve using 17 years of LAT data, by allowing the flux normalization of the sources within $3^{\circ}$ vary freely while freezing all other parameters of the analysis. We display the results in Fig.~\ref{fig:lc}, with 1-year time binning (last point has a bin size smaller than 1 year), showing that the unID flux has remained flat over the entire period under consideration. In particular, the variability index obtained in the analysis is $\mathrm{TS}_\textit{var}\sim 20.64$. We recall that, according to the 4FGL-DR4 catalog and the number of time intervals considered here (N=18), a source would be variable if $\mathrm{TS}_\textit{var}$ $>$ 33.4 \footnotemark[3].

\begin{figure}[h]
\begin{center}
        {\includegraphics[height=0.7\linewidth]{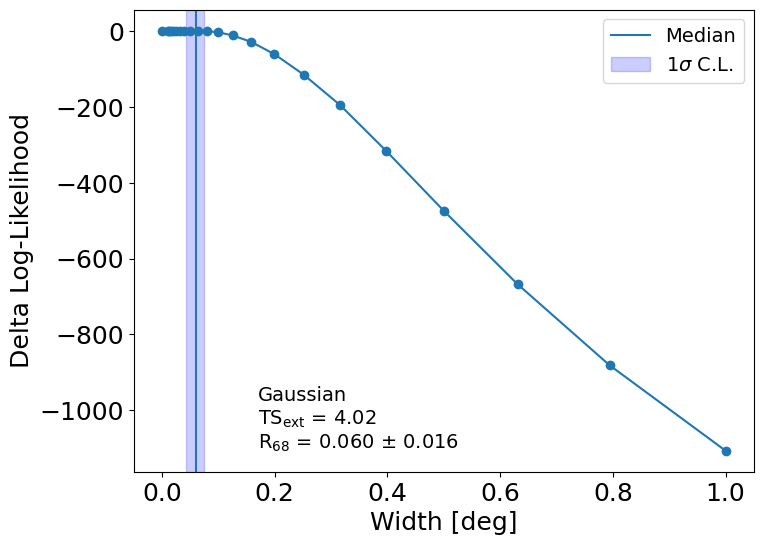}}\\
        {\includegraphics[height=0.7\linewidth]{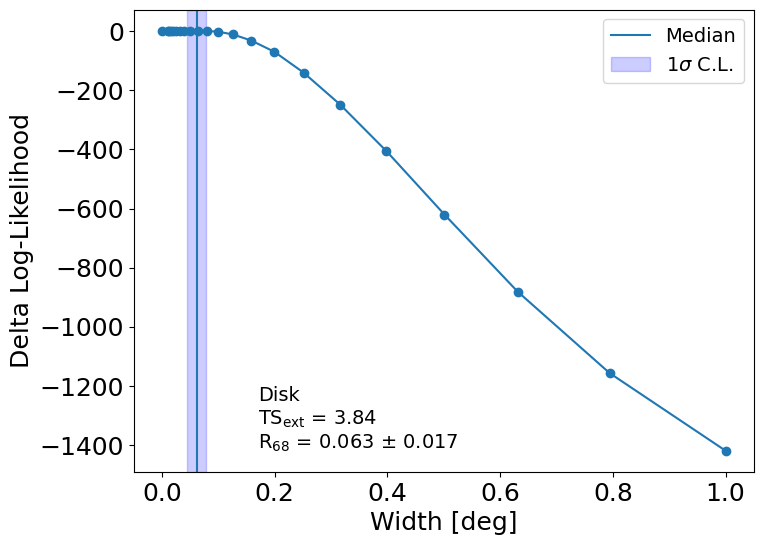}}
        \caption{Spatial extension log-likelihood profile of 4FGL J2112.5-3043, adopting a 2D Gaussian profile (top panel), and a 2D disk profile (bottom panel). The blue vertical line, along with the 1$\sigma$ uncertainty band, represents the median best-fit value.
        }
\label{fig:spatial}
\end{center}
\end{figure}

Finally, we perform a spatial analysis of our unID. According to several studies based on $N$-body cosmological simulations, spatial extension of a source is considered a hint for DM annihilation (e.g., ~\cite{Bertoni:2016hoh, chou2017resolvingdarkmattersubhalos, Ackermann_2018}). In particular, the brightest of the dark satellites are expected to be extended objects, with total sizes of up to few degrees \cite{Aguirre-Santaella:2023sww}, although they should be seen as $\sim$ 0.3 degrees size objects by the LAT according to ref. \cite{Coronado_Bl_zquez_2022}. We run the \texttt{GTAnalysis.extension} module, in which the spatial extension is defined as the $68\%$ containment width of the signal. \emph{Fermipy} fits the source with one spatial profile in concentric circles of increasing radii, centered at the position of the source, and computes the likelihood value for each of them. We perform a TS analysis to account for the extended model preference over a point-like source model (TS$_{ext}$), following the definition in Eq. \ref{eq:ts}, where $\mathcal{L} (H_0)$ is now the point-like source model (null hypothesis) and $\mathcal{L} (H_1)$ is the extended source model (alternative hypothesis). We computed the spatial distribution using two different models, a 2D Gaussian and a 2D uniform disk. The top panel in Fig.~\ref{fig:spatial} shows the log-likelihood profile and corresponding TS$_{ext}$ values of the extension evaluated with the Gaussian profile, whereas the bottom one in Fig.~\ref{fig:spatial} shows the corresponding results obtained with the 2D disk model. In both cases, only a marginal spatial extension is obtained, with a 68\% containment width of
$(0.060 \pm 0.016)^\circ$ and $(0.063 \pm 0.017)^\circ$ for the 2D Gaussian and 2D disk templates, respectively.
However, these values are below the \emph{Fermi}-LAT Pass 8 point-spread function (PSF) in the energy range
considered, which for example has a $68\%$ containment radius of $\sim 0.1^\circ$ at $10$ GeV. Therefore, the measured extension is compatible with the instrumental PSF and cannot be regarded as significant.

%In both cases, no significant spatial extension is found, contrary to what would be expected for a dark subhalo, with a 68$\%$ containment width of $0.060 \pm 0.016$ and $0.063 \pm 0.017$ for the 2D Gaussian and the 2D disk templates, respectively. 

\section{Possible Origin Scenarios for 4FGL J2112.5-3043}
\label{sec:intepretation}

The \emph{Fermi}-LAT covered energy range is well-suited to detecting gamma-ray annihilation or decay products of one of the most popular cold DM candidates, the so-called Weakly Interacting Massive Particles (WIMPs) (e.g. \cite{Bertone:2010at, Arcadi:2024ukq}).
%Such particles could reproduce the observed DM relic density~(\cite{Planck:2018vyg}) if we assume that they have a GeV-TeV mass scale (\cite{Bertone_2005, Bertone_2018, Bertone_2010}), weak-scale couplings, were produced in the early Universe, and subsequently decoupled from the plasma under thermal equilibrium conditions~\cite{Gondolo:1990dk}. WIMP self-annihilation can produce a photon flux with different contributions: broad continuum gamma-ray spectra from decay and fragmentation of the Standard Model (SM) particles produced in all annihilation processes (e.g. $b\bar{b}$,\,$\tau^+\tau^-$,\,$ WW$) , or sharp spectral features from virtual internal bremsstrahlung (VIB), as well as loop-induced processes ($\chi\chi \rightarrow \gamma\gamma,\,\gamma Z,\,\gamma H$). These signatures, expected in different WIMP scenarios including neutralinos in supersymmetric models \cite{Goldberg:1983nd, Ellis:1983ew, Jungman:1995df} or simplified portal models \cite{Abdallah:2014hon, Abdallah:2015ter, Arina:2016cqj, Arina:2025zpi}, make DM subhalos in the Galactic halo very compelling targets. 
The expected high DM density of dark satellites could yield observable gamma-ray signals in the \emph{Fermi}-LAT energy window. UnID \emph{Fermi}-LAT sources thus provide an ideal laboratory to test these WIMP models. Indeed, a source emitting only at gamma-ray energies with no multiwavelength counterpart and exhibiting spatial extension, would constitute a hint for DM annihilation. In this context, 4FGL J2112.5-3043 displays several characteristics consistent with this scenario: it shows no obvious multiwavelength counterparts within twice its $95\%$ C.L. error ellipse, neither a significant variability over the entire 17 years of \textit{Fermi}-LAT data. However, this source is consistent with a point-like spatial morphology, contrary to what we would expect for a dark satellite. As for its spectrum, our LAT analysis shows that this unID has a SED that is well described by an LP model, which is valid for describing both a pulsar and a DM source.

%In this context, 4FGL J2112.5-3043 displays several characteristics consistent with this scenario: it shows no obvious multiwavelength counterparts within twice its $95\%$ C.L. error ellipse, neither a significant variability over the entire 17 years of \textit{Fermi}-LAT data. However, it is consistent with a point-like morphology, contrary to what we would expect for a DM subhalo. On the other hand, our LAT analysis shows that this unID has a SED that is well described by an LP model, which is valid for describing both a pulsar and a DM spectrum.    

Indeed, in the context of DM searches with gamma rays, there has been confusion in disentangling between the typical spectra of pulsars from those that could be due to DM annihilation signals (e.g.,~\cite{Mirabal:2016huj, Fermi-LAT:2012fij}), especially for low WIMP DM masses and particular annihilation channels. Ref.~\cite{Coronado-Blazquez:2019pny} introduces a novel test based on the so-called ``$\beta$-plot'', to help elucidate between these objects. %The latter plot represents the curvature spectral index, $\beta$, of the sources versus their peak energy, $E_{peak}$.
In particular, their original test focused on those sources from the 4FGL \textit{Fermi}-LAT source catalog \cite{Fermi-LAT:2019yla} whose spectral best-fit is given by a log-parabola as in Eq. \ref{equation:LP}.

%where $K$ is the gamma-ray flux normalization, $E_0$ is the pivot energy that sets the energy scale, and $\alpha$ is the photon spectral index. The curvature spectral index, $\beta$, is also included. %This function fits well blazars and pulsars, although it could also be used to fit DM spectra in some cases. 
%\cfs{The LP function is already introduced in section \ref{sec:data analysis}. We should decide where we want to keep it.}\masc{good point! I'd say first time we use it, sorry I didn't see this before. If this is already in sec3, refer here properly to that, including equation number, etc.}

The $\beta$-plot represents the curvature spectral index, $\beta$, of  sources versus their peak energy, $E_{\rm peak}$. The latter refers to the energy at which the maximum energy flux is reached:
\begin{equation}
    E_{\rm peak} = E_{0} e^{\frac{2-\alpha}{2\beta}}.
    \label{eq:LP}
\end{equation}

\begin{figure}[h]
\begin{center}
        {\includegraphics[height=0.75\linewidth]{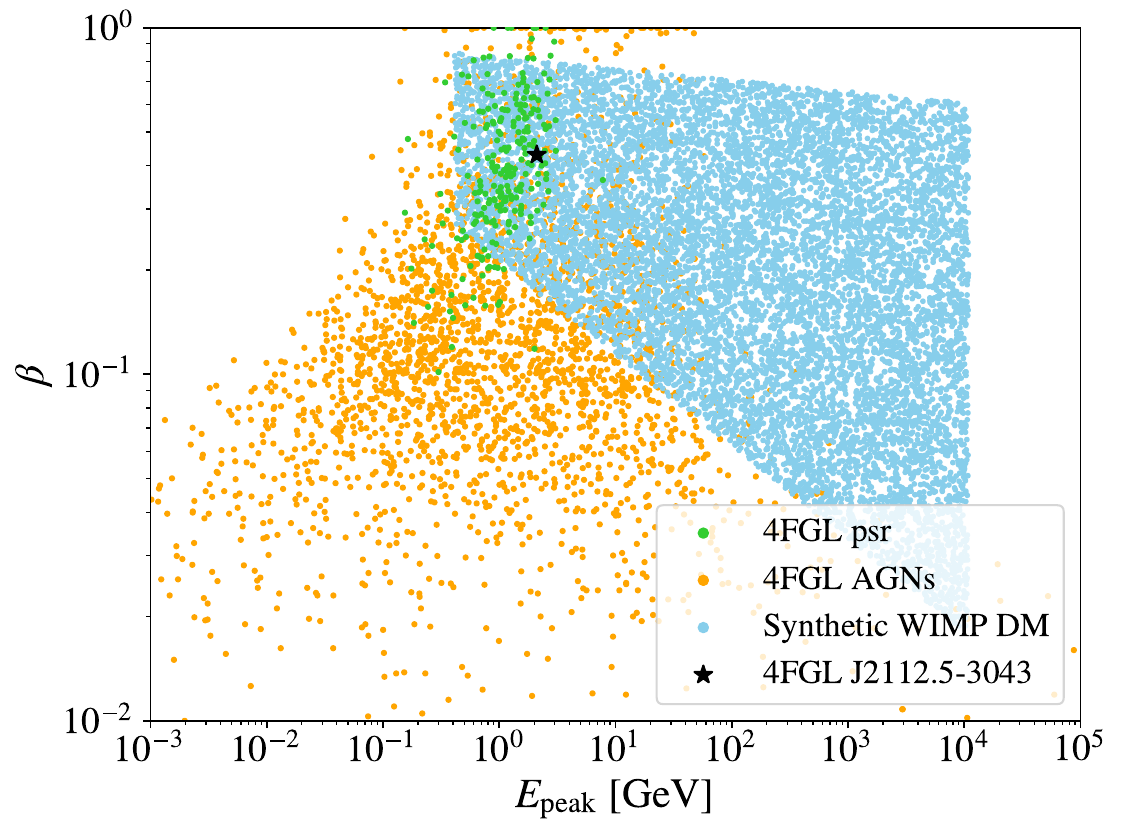}}
        \caption{Spectral curvature index ($\beta$) vs. the peak energy ($E_{\rm peak}$) — the so-called $\beta$-plot —, for all pulsars and AGNs present in the 4FGL \textit{Fermi} point-source catalog \cite{Fermi-LAT:2019yla}. Green points represent pulsars (``psr''), while orange points refer to AGNs. The blue points correspond to the synthetic WIMP DM data set in ref.~\cite{Gammaldi:2022wwz}. In this plot, 4FGL J2112.5-3043 is represented by a black star and lies in the pulsar/DM overlapping region.}
\label{fig:parameterspace}
\end{center}
\end{figure}

Figure~\ref{fig:parameterspace} shows the position, in the $\beta$-plot, of all pulsars (green points) and active galactic nuclei (AGNs, orange points) present in the 4FGL catalog. Shown in the same plot there is also a ``cloud'' of DM points (blue points), that were generated artificially in ref.~\cite{Gammaldi:2022wwz} for WIMP masses between 5 GeV $-$ 10 TeV and the $b\bar{b}$, $c\bar{c}$, $t\bar{t}$, $\tau\tau^{-}$, $e^+ e^-$, $\mu^+\mu^-$, $W^+W^-$, $ZZ
$, and $hh$  annihilation channels. In all these cases, the procedure consisted in producing the DM annihilation spectrum for each mass and channel considered, and fit it with an LP to derive the corresponding pair of $E_{\rm peak}$ and $\beta$ values. As it can be seen, there is an overlap between some of the synthetic DM points and the actual astrophysical sources, mainly pulsars. This happens especially for low WIMP masses, and for the $b\bar{b}$ and $c\bar{c}$ annihilation channels; see the discussion in ref.~\cite{Coronado-Blazquez:2019pny}. 

We place 4FGL J2112.5-3043, our unID of interest, in the $\beta$-plot of Fig.~\ref{fig:parameterspace}; with values $\beta=0.43$ and $E_{\rm peak}=2.11$ GeV. Should the source be a candidate for DM, it would lie in the DM region. Unfortunately, and perhaps not surprisingly, 4FGL J2112.5-3043 is located in the $E_{\rm peak}-\beta$ area of the plot where the pulsar/DM confusion resides, making this test useful but not conclusive in this specific case.

\subsection{Astrophysical interpretation: pulsar origin}
\label{subsec:pulsar}

From the $\beta$-plot test we performed, we see that our unID is indeed very intriguing, not being able to discern a preference towards a DM or a pulsar origin. However, given its spectral shape, the unID 4FGL J2112.5-3043 is a strong pulsar candidate. Previous works (\cite{Clark:2022tjf, Wu:2017mnz}) have carried out analyses to investigate the pulsar nature of some candidates, including our source.
No significant pulsations have been detected from this source, either in radio or in gamma rays~\cite{Clark:2022tjf}. Nevertheless, several astrophysical interpretations remain plausible, including a black widow system, a binary pulsar, or a neutron star with a small spin period \cite{Pathania:2025iyj}. In particular, the MSPs exhibit more stable, short-period pulsations compared to ordinary pulsars, are characterized by very small spin-down rates, and are most commonly found in binary systems. They are believed to form through the recycling of old, slowly rotating neutron stars that have ceased pulsar activity. During accretion from a low-mass companion, both mass and orbital angular momentum are transferred to the neutron star, spinning it up to millisecond periods and reactivating its pulsar emission \cite{Manchester:2017azn}. Due to their stable and relatively high gamma-ray luminosities, MSPs can be detected in principle at extragalactic distances as well \cite{crawford2005search}. Since 4FGL J2112.5-3043 is compatible with an extragalactic origin, MSPs therefore represent a viable candidate for its identification.

%\fg{Make a comparison with some MSPs as sample..which one?}
%\apg{If I understood you used ALL MSPs meaured by Fermi in some recent calculations?}

In the third \textit{Fermi}-LAT pulsar catalog~\citep{Fermi-LAT:2023zzt} the classical subexponential cutoff power-law (PLEC) model was superseded by the more flexible PLEC4 parameterization, introduced in 4FGL-DR3 (see ref.~\citep{Fermi-LAT:2023zzt} for details):
\begin{equation}
    \frac{dN}{dE}=N_0\Bigg(\frac{E}{E_0}\Bigg)^{-\Gamma+d/b}\text{Exp}{\Bigg[\frac{d}{b^2}\Bigg(1-\frac{E}{E_0}\Bigg)^b\Bigg]},
    \label{eq:PLEC4}
\end{equation}
with $N_0$ being the flux density at $E_0$, $\Gamma$ is the local spectral index at $E_0$, $d$ is the local curvature at $E_0$, and $b$ is the exponential cutoff steepness parameter.

%LogLike(pulsar)$=56138.108869669086$

%TS(pulsar)$=4743.741171085974$

To test whether our source is compatible with a pulsar origin, we adopted the PLEC4 model to characterize its spectrum, as it provides a good description of the typical spectral shape of pulsars. Next, we perform a log-likelihood comparison between astrophysical models using the Akaike Information Criterion (AIC)~\cite{Akaike:1974vps}. The AIC is given by:
\begin{equation}
    \text{AIC}=2 k-2\text{log}(\mathcal{L}_{max}),
    \label{AIC}
\end{equation}
where $\mathcal{L}_{max}$ is the maximum likelihood and $k$ is the number of degrees of freedom (d.o.f.)\footnote{Models with higher d.o.f. may provide better fits, so to compare the log-likelihoods of different models we take into account their d.o.f. by applying the AIC method. However, we note that these models are not nested, which may have a significant impact in the precise interpretation of our results (see, e.g., ref. \citep{2016MNRAS.458L..84A}).}. In particular, a power-law (PL) model has $k=3$, while the LP model has $k=4$ and a power-law with super-exponential cutoff (PLEC) model has $k=5$. %\textcolor{blue}{This method indicates which model is preferred among the considered scenarios and it cannot be interpreted as a statistical significance \cite{BurnhamAnderson2004}}. 
%The $\Delta$AIC method features which model is preferred among the considered scenarios:
Therefore:
\begin{equation}
\Delta\mathrm{AIC}=\mathrm{AIC}_i-\mathrm{AIC}_{0}= 2 \cdot \Bigg[(k_i-k_0)+\log\Bigg(\frac{\mathcal{L}_0}{\mathcal{L}_i}\Bigg)\Bigg].
\label{deltaAIC}
\end{equation}
The AIC method indicates which model is preferred among the considered scenarios. 
According to Eq. \ref{deltaAIC}, we compare the AIC value of the default model with the AIC value of the alternative model. Specifically, in the range $-2\leq \Delta \mathrm{AIC}\leq 0$ both models are comparable, $-7\leq \Delta \mathrm{AIC}\leq -4$ means that the alternative model is slightly preferred, while for $\Delta \mathrm{AIC}<-10$ there is a clear preference for the alternative model \cite{BurnhamAnderson2004}.
%\textcolor{blue}{In the range $0\leq \Delta \mathrm{AIC}\leq 2$ the models are comparable, $4\leq \Delta \mathrm{AIC}\leq 7$ means that the alternative is slightly preferred, and for $\Delta \mathrm{AIC}>10$ there is a clear preference for the alternative model \cite{BurnhamAnderson2004}.}
In particular, we compare the likelihood $\mathcal{L}_0$ of the best-fit astrophysical spectral model of the source given by the LAT analysis, i.e., the LP model, with the likelihood $\mathcal{L}_i$ corresponding to the alternative model, i.e., the PLEC4 model.
Performing the log-likelihood comparison via the AIC method between both astrophysical models, we obtain $\Delta \mathrm{AIC}=-1.06$ in favor of the PLEC4 model. Therefore, we found that, while both models adequately reproduce the spectrum of our unID, the PLEC4 model provides a slightly better fit. %to the spectrum of our unID.

\subsection{DM interpretation} 
\label{subsec:new physics}

%\masc{we should  work on this part of the text, too! it looks unfinished to me?}Hint for DM origin is the spatial extension target: galactic center (unresolved diffuse galactic emission with a big background), dwarf spheroidal galaxies (low background but faint DM signal and big uncertainties of DM model), and dark satellites means subhalos, completely dark and compact with no astrophysical background but big uncertainties in their structural distribution, number and sky position.

Based on our spectral and spatial analysis results, we cannot conclusively determine whether or not this source is compatible with DM.
%From the $\beta$-plot test we performed, we see that our unID is indeed very intriguing, not being able to discern a preference towards a DM or a pulsar origin. 
Here we carry out a more rigorous study to elucidate the origin of this source using the AIC introduced in Eq. \ref{deltaAIC}. Now, we compare the likelihood $\mathcal{L}_0$ of the best-fit astrophysical spectral model of the source with the likelihoods $\mathcal{L}_i$ corresponding to the predictions for %gamma-ray emission from 
WIMP annihilation into various final states, modeled with the \texttt{DMFitFunction}~\cite{Jeltema:2008hf}, where the DM via \texttt{DMFitFunction} has $k=2$.

%, as: 
%\begin{equation}
%    \Delta \mathrm{TS} = 2 \cdot \Bigg[(k_0-k_i)+\log\Bigg(\frac{\mathcal{L}_i}{\mathcal{L}_0}\Bigg)\Bigg],
%    \label{eq:DeltaTS}
%\end{equation}

%where $k_0$ and $k_i$ are the numbers of degrees of freedom (d.o.f.) of each respective model\footnote{Models with higher d.o.f. may provide better fits, so to compare the log-likelihoods of different models we take into account their d.o.f. by applying the AIC method. \textcolor{blue}{However, we note that these models are not nested, which may have a significant impact in the precise interpretation of our results (ref: ask Miguel!).}}. %with $k=3$ for a PL, $k=4$ for a LP and $k=5$ for a PLEC.
%In particular, a power-law (PL) model has $k=3$, while the \textcolor{blue}{DM via }\texttt{DMFitFunction} has $k=2$, LP has $k=4$ and a power-law with super-exponential cutoff (PLEC) model has $k=5$.

%[\textcolor{blue}{We have explored different annihilation channels, finding a preference for hadronic ones, while channels involving leptons or bosons yield negative $\Delta\text{TS}$ values and therefore show no preference for DM. Indeed,}] 
We have explored different annihilation channels ($b\bar{b}$, $c\bar{c}$, $\tau\tau^{-}$, $e^+ e^-$, and $\mu^+\mu^-$), finding a preference for the hadronic ones. We note that, considering the SED of 4FGL J2112.5-3043, with the last data point located at 20 GeV, the $W^{+}W^{-}$, $ZZ$, $t\bar{t}$ and $hh$ channels are cinematically forbidden. The comparison between the LP astrophysical model and DM provides a preference for a DM origin in the $b\bar{b}$ and $c\bar{c}$ annihilation channels, with $\Delta (\mathrm{AIC})=-13.37$ and $\Delta (\mathrm{AIC})=-12.31$, respectively. The similarity of the $\Delta (\mathrm{AIC})$ values for these two channels simply reflects the similarity of their spectral shapes, with both channels being equally viable. This result shows a qualitative preference for the DM hypothesis, yet it is not enough to robustly prefer DM against the conventional astrophysical scenario, due to the associated uncertainties. For instance, other DM annihilation spectra generators such as CosmiXs \cite{arina2023cosmixscosmicmessengerspectra} exist that would lead to slightly different DM spectra and, thus, to different AIC values. Moreover, if we were to account for the trial factors, we would expect the preference to decrease even further.
%In fact, this preference is expected to significantly decrease once corrected by the large number of trials employed here (i.e. fits to 5 different channels and $[5 - 50]$ GeV DM masses per channel).}

Additionally, as we discussed in Sec. \ref{subsec:pulsar}, the PLEC4 model provides a slightly better fit than the LP model. Then, we also make a comparison via AIC between PLEC4 and the DM hypothesis. As shown in Table \ref{tab:TSDMpulsar}, we obtain that the two favored DM annihilation channels ($b\bar{b}$ and $c\bar{c}$) also provide a better fit than the PLEC4 model, with $\Delta$(AIC) = -12.31 and $\Delta$(AIC) = -11.25, respectively.

\begin{table}[h!]
\begin{center}
\begin{tabular}{p{1.5cm} |p{1.5cm} |p{2cm}|p{2.5cm}}
\hline
\hline
      Channel   &  $\Delta (\mathrm{AIC})$ & $m_{DM}$ (GeV) & $\sigma v\,(10^{-26})\,\mathrm{cm}^{3} \mathrm{s}^{-1}$\\
       \hline
       \hline
      % $e^+e^-$  &  $-418.52$ & $9.99 \pm 2.37$ & $5.51\pm 2.01$\\
      % $\mu^+\mu^-$ &  $-170.38$ & $7.50 \pm 4.05$ & $6.37 \pm 4.36$\\
      % $\tau^+\tau^-$ &  $-140.79$ & $10.82 \pm 0.25$ & $0.94\pm 0.03$\\
       $b\overline{b}$ &  $-12.31$ & $46.2 \pm 1.4$ & $2.37\pm 0.10$\\
       %$W^+W^-$ & $10.16$ & $38.70 \pm 1.69$ & $2.64\pm 0.16$\\
       %$ZZ$ & $-139.84$ &  $20.49 \pm 0.49$ & $1.37\pm0.05$\\
       $c\overline{c}$ & $-11.25$ & $33.0 \pm 1.2 $ & $1.57\pm 0.08$\\
       \hline
    \end{tabular}
    \caption{$\Delta (\mathrm{AIC})$ values, as given by Eq. \ref{deltaAIC}, between the best-fit astrophysical model (PLEC4) and the best-fit values for the DM mass and the annihilation cross-section ($\sigma v$), for the $b\bar{b}$ and $c\bar{c}$ channels. Here we only report the DM mass that maximizes the likelihood for each channel. Moreover, we only show the results for those channels providing positive $\Delta \mathrm{AIC}$ values, and thus having a preference for the DM scenario.} 
    \label{tab:TSDMpulsar}
    \end{center}
\end{table}

\begin{figure}[h]
    \centering
    \includegraphics[width=1\linewidth]{spectrum_bb.png}
    \caption{The gamma-ray energy spectrum of 4FGL 2112.5-3043, where the black curve denotes the predicted fit to the $b\bar{b}$ annihilation channel with a DM mass of $(46.2 \pm 1.4)$ GeV.}
    % dark matter particle that annihilates to $b\bar{b}$
    \label{fig:bbbar}
\end{figure}

Figure \ref{fig:bbbar} displays the energy spectrum of 4FGL 2112.5-3043, together with the best-fit assuming DM annihilating to $b\bar{b}$ for a DM particle mass of $46.2$ GeV. 
%\fg{We could also add that the $t\bar{t}$ is not showed because this process is affected by IR divergencies and DMFitFuction does not eliminate them.} 
According to our results, if 4FGL J2112.5-3043 is indeed a dark satellite, it would imply a DM mass of $\sim33-46$ GeV with an annihilation cross-section of $\sigma v \sim 2\times 10^{-26} \mathrm{cm}^{3} \mathrm{s}^{-1}$, for the preferred cases of DM annihilation to $b\bar{b}$ and $c\bar{c}$.
%Note that considering the SED of 4FGL J2112.5-3043, with the last data point located at 20 GeV, the $W^{+}W^{-}$, $ZZ$, $t\bar{t}$ and $hh$ channels are cinematically forbidden.
%\textcolor{blue}{The values of the cross-section in Table \ref{tab:TSDMpulsar} are in tension with the limits given in Fig.3 of ref \citep{DelaTorreLuque:2024ozf} at least for the standard NFW profile.}
Note, however, that the required cross-section values for the $b\bar{b}$ and $c\bar{c}$ channels (see Table \ref{tab:TSDMpulsar}) are in tension with the current DM constraints provided by cosmic-ray antiproton measurements with the Alpha Magnetic Spectrometer (AMS-02), at least for the standard Navarro-Frenk-White DM density profile \citep{DelaTorreLuque:2024ozf}, as well as with the current limits given by dSphs \citep{mcdaniel2023legacyanalysisdarkmatter}.

It is worth noting that massive DM could accumulate inside compact objects such as pulsars or white dwarfs. These highly magnetized and dense stars, with strong magnetic fields and electron-rich atmospheres, provide a hybrid scenario in which astrophysical bodies may act as sinks for annihilating DM \cite{cermenoPhysRevD.98.063002}. For sufficiently old pulsars, however, heavy annihilating DM candidates become less viable, as their presence could trigger a transition into a more compact object \cite{herreroPhysRevD.100.103019}.

%\begin{table}[h!]
%\begin{center}
%\begin{tabular}{p{1.5cm} |p{1.5cm} |p{2cm}|p{2.5cm}}
%\hline
%\hline
 %     Channel   &  $\Delta \mathrm{TS}$ & $m_{DM}$ (GeV) & $\sigma v\,(10^{-26})\,\mathrm{cm}^{3} \mathrm{s}^{-1}$\\
      % \hline
      % \hline
      % $e^+e^-$  &  $-418.52$ & $9.99 \pm 2.37$ & $5.51\pm 2.01$\\
      % $\mu^+\mu^-$ &  $-170.38$ & $7.50 \pm 4.05$ & $6.37 \pm 4.36$\\
      % $\tau^+\tau^-$ &  $-140.79$ & $10.82 \pm 0.25$ & $0.94\pm 0.03$\\
      % $b\overline{b}$ &  $13.37$ & $46.2 \pm 1.4$ & $2.37\pm 0.10$\\
       %$W^+W^-$ & $10.16$ & $38.70 \pm 1.69$ & $2.64\pm 0.16$\\
       %$ZZ$ & $-139.84$ &  $20.49 \pm 0.49$ & $1.37\pm0.05$\\
      % $c\overline{c}$ & $12.31$ & $33.0 \pm 1.2 $ & $1.57\pm 0.08$\\
      %\hline
    %\end{tabular}
    %\caption{$\Delta \mathrm{TS}$ values, \textcolor{blue}{as given by Eq.\ref{eq:DeltaTS} between the best-fit astrophysical (LP) model and the} annihilation cross-section ($\sigma v$), for \textcolor{blue}{the} $b\bar{b}$ and $c\bar{c}$ channels. For clarity, here we reports only the DM mass that maximizes $\Delta \mathrm{TS}$ for each channel.} 
   % \label{tab:TSDM}
    %\end{center}
%\end{table}

%\fg{PUT STELLAR STREAM INTERPRETATION?} \cfs{I think it's best to explain this only in the introduction and perhaps also comment on it in the conclusions, since here it would involve 'breaking' a bit the story we are telling with our results.}

\section{Conclusions}

%In conclusion, i
In this work, we have deeply studied the unidentified gamma-ray source 4FGL J2112.5-3043 with the aim of shedding light on its possible nature. Among the plethora of unassociated sources in the latest \textit{Fermi}-LAT point-source catalog (4FGL-DR4), this source appears as the unID with the highest detection significance. After a detailed multiwavelength search, we found no potential association with this source, further reinforcing its intriguing origin.

We have studied the spectrum, variability, and spatial extension of its emission using 17 years of \textit{Fermi}-LAT data in the energy range between 100 MeV and 1 TeV. The main results of the analysis can be summarized as follows:
\begin{itemize}
    \item We have found source detection in the energy range from $\sim$ 300 MeV to $\sim$ 20 GeV. According to the 4FGL-DR4 catalog, the gamma-ray spectrum of 4FGL J2112.5-3043, is well described by a log-parabola model (Eq. \ref{eq:LP}), typically used to fit pulsar spectra but also compatible with predictions for some DM annihilation channels. In fact, this source lies in the $\beta$-plot region where the pulsar and DM populations overlap, making this test informative but not conclusive.
    On the other hand, our analysis shows that a PLEC4 model (Eq. \ref{eq:PLEC4}) — the one providing the best description for typical pulsar spectra — is slightly preferred against the log-parabola model via an AIC analysis and, thus, may constitute the best astrophysical description of our unID.
    
    \item Using a time binning of one year, we found no significant flux variability over 17 years of LAT data. This is consistent with the steady emission expected from DM annihilation.
    \item The spatial morphology of the emission is compatible with a point-like source. We note that this result is in tension with expectations for a dark satellite, which should exhibit some degree of extension (e.g., ~\cite{Bertoni:2016hoh, chou2017resolvingdarkmattersubhalos, Ackermann_2018, Aguirre-Santaella:2023sww}). In particular, in the $\Lambda$CDM framework, the brightest dark satellites of our Galaxy should be seen as extended objects of about 0.3 degrees by the LAT (\citep{Coronado_Bl_zquez_2022}).
    %the point-like appearance may result from the limited angular resolution of the \textit{Fermi}-LAT, which might not resolve a small intrinsic extension, or it could indicate that the source is located at a large distance.
\end{itemize}

%As a matter of fact, a source emitting only at gamma-ray energies with no multiwavelength counterpart and exhibiting spatial extension, would constitute a hint for DM annihilation, since known spatially extended astrophysical sources are also bright at other wavelengths. Based on our analysis results, however, we cannot firmly conclude whether 4FGL J2112.5-3043 is a dark subhalo or its origin is purely astrophysical.

We performed a comparison between different spectral models for our unID using the AIC method — that is, comparing the best-fit astrophysical models (log-parabola and PLEC4) with fits to various DM annihilation channels — and found what seems to be a clear preference for DM annihilation to $b\bar{b}$ and $c\bar{c}$ channels (expected to decrease significantly if trial factors were to be applied). In particular, should 4FGL J2112.5-3043 be a dark satellite, it would imply a DM mass in the range $\sim33-46$ GeV with an annihilation cross-section of $\sigma v \sim 2\times 10^{-26} \mathrm{cm}^{3} \mathrm{s}^{-1}$. 
Yet, despite the strong statistical preference indicated by the $\Delta$AIC values, this result may still depend on specific modeling assumptions. In particular, testing alternative DM models, considering a different DM annihilation spectra generator such as CosmiXs \citep{arina2023cosmixscosmicmessengerspectra}, or a combination of multiple annihilation channels could affect the inferred preference. Independent confirmation through more comprehensive statistical and phenomenological analyses is therefore required before drawing firm conclusions.

%Yet, significant uncertainties remain (e.g. the precise shape of the DM annihilation spectrum itself) that may significantly impact the statistical analysis, preventing us from making any claims at this point.
%\cfs{We'd need to include caveats here!}

%On the other hand, the detection of a hundred-TeV neutrino neutrino near the location of our unID, if associated with this source, introduces additional uncertainty to the new physics interpretation. 

%\textcolor{blue}{However}, based on our analysis results, we cannot firmly conclude whether 4FGL J2112.5-3043 is a dark \textcolor{blue}{satellite} or its origin is purely astrophysical. In conclusion, 4FGL J2112.5-3043 remains an intriguing object with the potential to be a \textcolor{blue}{dark satellite}, although an astrophysical scenario remains \textcolor{blue}{equally} plausible. 

%Our results therefore motivate further investigation. %\textcolor{blue}{We suggest a systematic search for potential associated dwarf galaxies, as well as a dedicated campaign of radio observations around the position of 4FGL J2112.5-3043. Although the lack of radio emission and gamma-ray pulsation is consistent with the MSP scenario in a compact binary system, deeper radio, optical, or X-ray campaigns could provide very valuable information about this source.}

%We suggest a dedicated campaign of radio observations around the position of 4FGL J2112.5-3043, as well as a systematic search for potential associated dwarf galaxies.
In conclusion, despite our efforts, our analysis does not allow us to unambiguously determine the nature of 4FGL J2112.5-3043. Nevertheless, this source remains an enigmatic object, as both an astrophysical interpretation and one in terms of DM remain viable. These results motivate further investigation. Although the lack of radio emission and gamma-ray pulsations is consistent with the MSP hypothesis in a compact binary system, deeper radio, optical (photometric and stellar dynamics) and X-ray campaigns could provide complementary information. %\textcolor{blue}{Another possible way to distinguish between the MSP and DM scenarios would be through a dedicated multiwavelength study of secondary emission. In particular, inverse Compton and synchrotron radiation from electrons and positrons could provide complementary signatures, with different spectral properties expected in the two cases. This could be investigated in future work.}
Another possible way to distinguish between the MSP and DM scenarios would be through a dedicated study of the expected secondary emission. In particular, we leave for future work the analysis of potential inverse Compton emission and synchrotron emission, which we expect to differ depending on whether a source is an MSP or its origin is DM, thus providing us with an alternative way to distinguish between both hypothesis.
Such observations would help to distinguish between astrophysical and DM scenarios, while a systematic search for potential associated baryonic material (e.g., an ultra-faint dwarf galaxy or compact stellar system) could offer additional insight.

\label{sec:conclusions}

\section*{Acknowledgements}

We thank David A. Smith for interesting and stimulating discussions on pulsars, as well as Fernando Valenciano-Ruano for their help in the spectral DM analysis and its interpretation. We also thank Davide Serini and Pedro De La Torre Luque for their comments, which have significantly improved this manuscript.

The \textit{Fermi} LAT Collaboration acknowledges generous ongoing support from a number of agencies and institutes that have supported both the development and the operation of the LAT as well as scientific data analysis. These include the National Aeronautics and Space Administration and the Department of Energy in the United States, the Commissariat \`a l'Energie Atomique and the Centre National de la Recherche Scientifique / Institut National de Physique Nucl\'eaire et de Physique des Particules in France, the Agenzia Spaziale Italiana and the Istituto Nazionale di Fisica Nucleare in Italy, the Ministry of Education, Culture, Sports, Science and Technology (MEXT), High Energy Accelerator Research Organization (KEK) and Japan Aerospace Exploration Agency (JAXA) in Japan, and the K.~A.~Wallenberg Foundation, the Swedish Research Council and the Swedish National Space Board in Sweden. 
Additional support for science analysis during the operations phase is gratefully 
acknowledged from the Istituto Nazionale di Astrofisica in Italy and the Centre 
National d'\'Etudes Spatiales in France. This work performed in part under DOE 
Contract DE-AC02-76SF00515.
%INFN and ASI personnel performed in part under ASI-INFN Agreements No. 2021-43-HH.0

Supported by Italian Research Center on High Performance Computing Big Data and Quantum Computing (ICSC), project funded by European Union - NextGenerationEU - and National Recovery and Resilience Plan (NRRP) - Mission 4 Component 2 within the activities of Spoke 3 (Astrophysics and Cosmos Observations).

F. G. and MAPG acknowledge financial support by Junta de Castilla y León  project  SA101P24.
%The work of CFS and MASC was supported by the grants PID2021-125331NB-I00 and CEX2020-001007-S, both funded by MCIN/AEI/10.13039/501100011033 and by ``ERDF A way of making Europe''. They also acknowledges the MultiDark Network, ref. RED2022-134411-T.
The work of CFS was partially supported by Programa Investigo 2022 Comunidad de Madrid with ref. A113, funded by European Union - NextGenerationEU, and by the ``La Caixa'' Foundation (ID 100010434) and the European Union's Horizon~2020 research and innovation program under the Marie Skłodowska-Curie grant agreement No~847648, fellowship code LCF/BQ/PI21/11830030. 
The work of CFS and MASC was 
supported by the grants PID2024-155874NB-C21 and CEX2020-001007-S, both funded by MCIN/AEI/10.13039/501100011033 and by ``ERDF A way of making Europe''.
CFS, MASC and MAPG also acknowledge the MultiDark Network, ref. RED2022-134411-T. MAPG acknowledge  partial finantial support by MICIU project PID2022-137887NB-I00, Gravitational Wave Network (REDONGRA) Strategic Network RED2024-153735-E funded by MICIU/AEI/10.13039/501100011033 and COST Action COSMIC WISPers CA21106.
%Finally, we thank David A. Smith for interesting and stimulating discussions on pulsars, as well as Dario Gasparrini and Stefano Ciprini for multiwavelength analysis. We also thank Fernando Valenciano for helpful discussions.

Facilities: \textit{Fermi Gamma-ray Space Telescope} --- \textit{LAT} --- \textit{unID}
%\textit{RXTE} ---
%\textit{OVRO}  --- \textit{Tuorla} --- \textit{KVA} --- \textit{KAIT} --- \textit{CSS} --- \textit{CRTS} --- \textit{ASAS-SN}

%..........

\bibliographystyle{elsarticle-num} 
\bibliography{biblio}

@article{Braglia:2020jri,
    author = "Braglia, C. and Mignani, R. P. and Belfiore, A. and Marelli, M. and Israel, G. L. and Novara, G. and De Luca, A. and Tiengo, A. and Saz Parkinson, P. M.",
    title = "{A multiwavelength search for black widow and redback counterparts of candidate {\ensuremath{\gamma}}-ray millisecond pulsars}",
    eprint = "2007.00442",
    archivePrefix = "arXiv",
    primaryClass = "astro-ph.HE",
    doi = "10.1093/mnras/staa2339",
    journal = "Mon. Not. Roy. Astron. Soc.",
    volume = "497",
    number = "4",
    pages = "5364--5382",
    year = "2020"
}

@article{Fermi-LAT:2009ihh,
    author = "Atwood, W. B. and others",
    collaboration = "Fermi-LAT",
    title = "{The Large Area Telescope on the Fermi Gamma-ray Space Telescope Mission}",
    eprint = "0902.1089",
    archivePrefix = "arXiv",
    primaryClass = "astro-ph.IM",
    reportNumber = "SLAC-PUB-13620",
    doi = "10.1088/0004-637X/697/2/1071",
    journal = "Astrophys. J.",
    volume = "697",
    pages = "1071--1102",
    year = "2009"
}

@article{Agashe:2014yua,
  author       = {Agashe, Kaustubh and Cui, Yanou and Necib, Lina and Thaler, Jesse},
  title        = {(In)Direct Detection of Boosted Dark Matter},
  journal      = {JCAP},
  year         = {2014},
  month        = {Oct},
  number       = {10},
  pages        = {062},
  doi          = {10.1088/1475-7516/2014/10/062},
  eprint       = {1405.7370},
  archivePrefix= {arXiv},
  primaryClass = {hep-ph}
}

@article{Bell:2021zkr,
  author       = {Bell, Nicole F. and Dent, James B. and Dutta, Bhaskar and Ghosh, Sumit and Kumar, Jason and Newstead, Jayden L. and Shoemaker, Ian M.},
  title        = {Cosmic-ray upscattered inelastic dark matter},
  journal      = {Phys. Rev. D},
  year         = {2021},
  volume       = {104},
  number       = {7},
  pages        = {076020},
  doi          = {10.1103/PhysRevD.104.076020},
  eprint       = {2108.00583},
  archivePrefix= {arXiv},
  primaryClass = {hep-ph}
}

@article{Clark:2022tjf,
    author = "Clark, C. J. and others",
    title = "{The TRAPUM L-band survey for pulsars in Fermi-LAT gamma-ray sources}",
    eprint = "2212.08528",
    archivePrefix = "arXiv",
    primaryClass = "astro-ph.HE",
    doi = "10.1093/mnras/stac3742",
    journal = "Mon. Not. Roy. Astron. Soc.",
    volume = "519",
    number = "4",
    pages = "5590--5606",
    year = "2023"
}

@article{Gehrels:1999ri,
    author = "Gehrels, N. and Michelson, P.",
    editor = "Catanese, M. and Weekes, T.",
    title = "{GLAST: The next generation high-energy gamma-ray astronomy mission}",
    reportNumber = "SLAC-REPRINT-1999-138",
    doi = "10.1016/S0927-6505(99)00066-3",
    journal = "Astropart. Phys.",
    volume = "11",
    pages = "277--282",
    year = "1999"
}

@article{Fernandez-Suarez_2025,
doi = {10.1088/1475-7516/2025/09/003},
url = {https://doi.org/10.1088/1475-7516/2025/09/003},
year = {2025},
month = {sep},
publisher = {IOP Publishing},
volume = {2025},
number = {09},
pages = {003},
author = {Fernández-Suárez, Cristina and Sánchez-Conde, Miguel A.},
title = {A search for dark matter annihilation in stellar streams with the Fermi-LAT},
journal = {Journal of Cosmology and Astroparticle Physics},
}

@article{Bertoni:2016hoh,
    author = "Bertoni, Bridget and Hooper, Dan and Linden, Tim",
    title = "{Is The Gamma-Ray Source 3FGL J2212.5+0703 A Dark Matter Subhalo?}",
    eprint = "1602.07303",
    archivePrefix = "arXiv",
    primaryClass = "astro-ph.HE",
    reportNumber = "FERMILAB-PUB-15-411-A, INT-PUB-16-005",
    doi = "10.1088/1475-7516/2016/05/049",
    journal = "JCAP",
    volume = "05",
    pages = "049",
    year = "2016"
}

@article{Balazs:2024uyj,
    author = "Balazs, Csaba and Bringmann, Torsten and Kahlhoefer, Felix and White, Martin",
    title = "{A Primer on Dark Matter}",
    eprint = "2411.05062",
    archivePrefix = "arXiv",
    primaryClass = "astro-ph.CO",
    month = "11",
    year = "2024"
}

@article{Massaro:2012pi,
    author = "Massaro, F. and D'Abrusco, R. and Tosti, G. and Ajello, M. and Paggi, A. and Gasparrini, D.",
    title = "{Unidentifed gamma-ray sources: hunting gamma-ray blazars}",
    eprint = "1203.3801",
    archivePrefix = "arXiv",
    primaryClass = "astro-ph.HE",
    reportNumber = "SLAC-PUB-14895",
    doi = "10.1088/0004-637X/752/1/61",
    journal = "Astrophys. J.",
    volume = "752",
    pages = "61",
    year = "2012"
}

@article{Wu:2017mnz,
    author = "Wu, J. and others",
    title = "{The Einstein@Home Gamma-ray Pulsar Survey. II. Source Selection, Spectral Analysis, and Multiwavelength Follow-up}",
    eprint = "1712.05395",
    archivePrefix = "arXiv",
    primaryClass = "astro-ph.HE",
    doi = "10.3847/1538-4357/aaa411",
    journal = "Astrophys. J.",
    volume = "854",
    number = "2",
    pages = "99",
    year = "2018"
}

@article{Coronado-Blazquez:2019pny,
    author = "Coronado-Bl{\'a}zquez, Javier and S{\'a}nchez-Conde, Miguel A. and Di Mauro, Mattia and Aguirre-Santaella, Alejandra and Ciuc{\u{a}}, Ioana and Dom{\'\i}nguez, Alberto and Kawata, Daisuke and Mirabal, N{\'e}stor",
    title = "{Spectral and spatial analysis of the dark matter subhalo candidates among $Fermi$ Large Area Telescope unidentified sources}",
    eprint = "1910.14429",
    archivePrefix = "arXiv",
    primaryClass = "astro-ph.HE",
    doi = "10.1088/1475-7516/2019/11/045",
    journal = "JCAP",
    volume = "11",
    pages = "045",
    year = "2019"
}

@article{herreroPhysRevD.100.103019,
  title = {Dark matter and bubble nucleation in old neutron stars},
  author = {Herrero, A. and P\'erez-Garc\'{\i}a, M. A. and Silk, J. and Albertus, C.},
  journal = {Phys. Rev. D},
  volume = {100},
  issue = {10},
  pages = {103019},
  numpages = {9},
  year = {2019},
  month = {Nov},
  publisher = {American Physical Society},
  doi = {10.1103/PhysRevD.100.103019},
  url = {https://link.aps.org/doi/10.1103/PhysRevD.100.103019}
}

@article{cermenoPhysRevD.98.063002,
  title = {Gamma rays from dark mediators in white dwarfs},
  author = {Cerme\~no, M. and P\'erez-Garc\'{\i}a, M. A.},
  journal = {Phys. Rev. D},
  volume = {98},
  issue = {6},
  pages = {063002},
  numpages = {9},
  year = {2018},
  month = {Sep},
  publisher = {American Physical Society},
  doi = {10.1103/PhysRevD.98.063002},
  url = {https://link.aps.org/doi/10.1103/PhysRevD.98.063002}
}

@article{Gammaldi:2022wwz,
    author = "Gammaldi, V. and Zald{\'\i}var, B. and S{\'a}nchez-Conde, M. A. and Coronado-Bl{\'a}zquez, J.",
    title = "{A search for dark matter among Fermi-LAT unidentified sources with systematic features in machine learning}",
    eprint = "2207.09307",
    archivePrefix = "arXiv",
    primaryClass = "astro-ph.HE",
    doi = "10.1093/mnras/stad066",
    journal = "Mon. Not. Roy. Astron. Soc.",
    volume = "520",
    number = "1",
    pages = "1348--1361",
    year = "2023"
}

@article{Fermi-LAT:2019yla,
    author = "Abdollahi, S. and others",
    collaboration = "Fermi-LAT",
    title = "{$Fermi$ Large Area Telescope Fourth Source Catalog}",
    eprint = "1902.10045",
    archivePrefix = "arXiv",
    primaryClass = "astro-ph.HE",
    doi = "10.3847/1538-4365/ab6bcb",
    journal = "Astrophys. J. Suppl.",
    volume = "247",
    number = "1",
    pages = "33",
    year = "2020"
}

@article{Mirabal:2016huj,
    author = "Mirabal, N. and Charles, E. and Ferrara, E. C. and Gonthier, P. L. and Harding, A. K. and S{\'a}nchez-Conde, M. A. and Thompson, D. J.",
    title = "{3FGL Demographics Outside the Galactic Plane using Supervised Machine Learning: Pulsar and Dark Matter Subhalo Interpretations}",
    eprint = "1605.00711",
    archivePrefix = "arXiv",
    primaryClass = "astro-ph.HE",
    doi = "10.3847/0004-637X/825/1/69",
    journal = "Astrophys. J.",
    volume = "825",
    number = "1",
    pages = "69",
    year = "2016"
}

@article{Fermi-LAT:2012fij,
    author = "Ackermann, M. and others",
    collaboration = "Fermi-LAT",
    title = "{Search for Dark Matter Satellites using the FERMI-LAT}",
    eprint = "1201.2691",
    archivePrefix = "arXiv",
    primaryClass = "astro-ph.HE",
    reportNumber = "SLAC-PUB-14974",
    doi = "10.1088/0004-637X/747/2/121",
    journal = "Astrophys. J.",
    volume = "747",
    pages = "121",
    year = "2012"
}

@article{Ballet:2023qzs,
    author = "Ballet, J. and Bruel, P. and Burnett, T. H. and Lott, B.",
    collaboration = "Fermi-LAT",
    title = "{Fermi Large Area Telescope Fourth Source Catalog Data Release 4 (4FGL-DR4)}",
    eprint = "2307.12546",
    archivePrefix = "arXiv",
    primaryClass = "astro-ph.HE",
    month = "7",
    year = "2023"
}

@inproceedings{Fermi-LAT:2013jgq,
    author = "Atwood, W. and others",
    collaboration = "Fermi-LAT",
    title = "{Pass 8: Toward the Full Realization of the Fermi-LAT Scientific Potential}",
    eprint = "1303.3514",
    archivePrefix = "arXiv",
    primaryClass = "astro-ph.IM",
    month = "3",
    year = "2013"
}

@article{Coronado-Blazquez:2019puc,
    author = "Coronado-Blazquez, Javier and Sanchez-Conde, Miguel A. and Dominguez, Alberto and Aguirre-Santaella, Alejandra and Di Mauro, Mattia and Mirabal, Nestor and Nieto, Daniel and Charles, Eric",
    title = "{Unidentified Gamma-ray Sources as Targets for Indirect Dark Matter Detection with the Fermi-Large Area Telescope}",
    eprint = "1906.11896",
    archivePrefix = "arXiv",
    primaryClass = "astro-ph.HE",
    doi = "10.1088/1475-7516/2019/07/020",
    journal = "JCAP",
    volume = "07",
    pages = "020",
    year = "2019"
}

@article{Pena-Herazo:2017evn,
    author = "Pe{\~n}a-Herazo, H. A. and others",
    title = "{Optical Spectroscopic Observations of Gamma-Ray Blazar Candidates. VII. Follow-up Campaign in the Southern Hemisphere}",
    eprint = "1903.10014",
    archivePrefix = "arXiv",
    primaryClass = "astro-ph.GA",
    doi = "10.1007/s10509-017-3208-7",
    journal = "Astrophys. Space Sci.",
    volume = "362",
    pages = "228",
    year = "2017"
}

@article{IceCube:2023agq,
    author = "Abbasi, R. and others",
    collaboration = "IceCube",
    title = "{IceCat-1: The IceCube Event Catalog of Alert Tracks}",
    eprint = "2304.01174",
    archivePrefix = "arXiv",
    primaryClass = "astro-ph.HE",
    doi = "10.3847/1538-4365/acfa95",
    journal = "Astrophys. J. Suppl.",
    volume = "269",
    number = "1",
    pages = "25",
    year = "2023"
}

@article{Wood:2017yyb,
    author = "Wood, Matthew and Caputo, Regina and Charles, Eric and Di Mauro, Mattia and Magill, Jeffrey and Perkins, Jeremy",
    collaboration = "Fermi-LAT",
    title = "{Fermipy: An open-source Python package for analysis of Fermi-LAT Data}",
    eprint = "1707.09551",
    archivePrefix = "arXiv",
    primaryClass = "astro-ph.IM",
    doi = "10.22323/1.301.0824",
    journal = "PoS",
    volume = "ICRC2017",
    pages = "824",
    year = "2018"
}

@article{Mirabal:2021ayb,
    author = "Mirabal, Nestor and Bonaca, Ana",
    title = "{Machine-learned dark matter subhalo candidates in the 4FGL-DR2: search for the perturber of the GD-1 stream}",
    eprint = "2105.12131",
    archivePrefix = "arXiv",
    primaryClass = "astro-ph.HE",
    doi = "10.1088/1475-7516/2021/11/033",
    journal = "JCAP",
    volume = "11",
    pages = "033",
    year = "2021"
}

@article{Fermi-LAT:2015att,
    author = "Ackermann, M. and others",
    collaboration = "Fermi-LAT",
    title = "{Searching for Dark Matter Annihilation from Milky Way Dwarf Spheroidal Galaxies with Six Years of Fermi Large Area Telescope Data}",
    eprint = "1503.02641",
    archivePrefix = "arXiv",
    primaryClass = "astro-ph.HE",
    reportNumber = "FERMILAB-PUB-15-081-AE",
    doi = "10.1103/PhysRevLett.115.231301",
    journal = "Phys. Rev. Lett.",
    volume = "115",
    number = "23",
    pages = "231301",
    year = "2015"
}

@article{Salvetti:2017pwl,
    author = {Salvetti, D. and Mignani, R. P. and De Luca, A. and Marelli, M. and Pallanca, C. and Breeveld, A. A. and Belfiore, A. and Becker, W. and Greiner, J. and H{\"u}semann, P.},
    title = "{A multiwavelength investigation of candidate millisecond pulsars in unassociated {\ensuremath{\gamma}}-ray sources}",
    eprint = "1702.00474",
    archivePrefix = "arXiv",
    primaryClass = "astro-ph.HE",
    doi = "10.1093/mnras/stx1247",
    journal = "Mon. Not. Roy. Astron. Soc.",
    volume = "470",
    number = "1",
    pages = "466--480",
    year = "2017"
}

@article{Arcadi:2024ukq,
    author = "Arcadi, Giorgio and Cabo-Almeida, David and Dutra, Ma{\'\i}ra and Ghosh, Pradipta and Lindner, Manfred and Mambrini, Yann and Neto, Jacinto P. and Pierre, Mathias and Profumo, Stefano and Queiroz, Farinaldo S.",
    title = "{The Waning of the WIMP: Endgame?}",
    eprint = "2403.15860",
    archivePrefix = "arXiv",
    primaryClass = "hep-ph",
    doi = "10.1140/epjc/s10052-024-13672-y",
    journal = "Eur. Phys. J. C",
    volume = "85",
    number = "2",
    pages = "152",
    year = "2025"
}

@article{Bertone:2010at,
    author = "Bertone, Gianfranco",
    title = "{The moment of truth for WIMP Dark Matter}",
    eprint = "1011.3532",
    archivePrefix = "arXiv",
    primaryClass = "astro-ph.CO",
    doi = "10.1038/nature09509",
    journal = "Nature",
    volume = "468",
    pages = "389--393",
    year = "2010"
}

@article{SWIFT:2005ngz,
    author = "Burrows, David N. and others",
    collaboration = "SWIFT",
    title = "{The Swift X-ray Telescope}",
    eprint = "astro-ph/0508071",
    archivePrefix = "arXiv",
    doi = "10.1007/s11214-005-5097-2",
    journal = "Space Sci. Rev.",
    volume = "120",
    pages = "165",
    year = "2005"
}

@article{Struder:2001bh,
    author = "Struder, L. and others",
    title = "{The European Photon Imaging Camera on XMM-Newton: The pn-CCD camera}",
    doi = "10.1051/0004-6361:20000066",
    journal = "Astron. Astrophys.",
    volume = "365",
    pages = "L18--26",
    year = "2001"
}

@article{Turner:2000jy,
    author = "Turner, Martin J. L. and others",
    title = "{The European Photon Imaging Camera on XMM-Newton: The MOS cameras}",
    eprint = "astro-ph/0011498",
    archivePrefix = "arXiv",
    doi = "10.1051/0004-6361:20000087",
    journal = "Astron. Astrophys.",
    volume = "365",
    pages = "L27--35",
    year = "2001"
}

@article{Elvis:2009rr,
    author = "Elvis, Martin and others",
    title = "{The Chandra COSMOS Survey, I: Overview and Point Source Catalog}",
    eprint = "0903.2062",
    archivePrefix = "arXiv",
    primaryClass = "astro-ph.CO",
    doi = "10.1088/0067-0049/184/1/158",
    journal = "Astrophys. J. Suppl.",
    volume = "184",
    pages = "158--171",
    year = "2009"
}

@article{Drake:2008vy,
    author = "Drake, A. J. and others",
    title = "{First Results from the Catalina Real-time Transient Survey}",
    eprint = "0809.1394",
    archivePrefix = "arXiv",
    primaryClass = "astro-ph",
    doi = "10.1088/0004-637X/696/1/870",
    journal = "Astrophys. J.",
    volume = "696",
    pages = "870--884",
    year = "2009"
}

@ARTICLE{GaiaDR3,
       author = {{Gaia Collaboration} and Vallenari, A. and Brown, A. G. A. and Prusti, T. and others},
       title = "{Gaia Data Release 3: Summary of the content and survey properties}",
       journal = {Astronomy \& Astrophysics},
       volume = {674},
       year = {2023},
       pages = {A1},
       doi = {10.1051/0004-6361/202243940}
}

@article{Jeltema:2008hf,
    author = "Jeltema, Tesla E. and Profumo, Stefano",
    title = "{Fitting the Gamma-Ray Spectrum from Dark Matter with DMFIT: GLAST and the Galactic Center Region}",
    eprint = "0808.2641",
    archivePrefix = "arXiv",
    primaryClass = "astro-ph",
    doi = "10.1088/1475-7516/2008/11/003",
    journal = "JCAP",
    volume = "11",
    pages = "003",
    year = "2008"
}

@article{Akaike:1974vps,
    author = "Akaike, H.",
    title = "{A new look at the statistical model identification}",
    doi = "10.1109/TAC.1974.1100705",
    journal = "IEEE Trans. Automatic Control",
    volume = "19",
    number = "6",
    pages = "716--723",
    year = "1974"
}

@article{Fermi-LAT:2025gei,
    author = "Abdollahi, S. and others",
    collaboration = "Fermi-LAT, HAWC, H.E.S.S., MAGIC, VERITAS",
    title = "{Combined dark matter search towards dwarf spheroidal galaxies with Fermi-LAT, HAWC, H.E.S.S., MAGIC, and VERITAS}",
    eprint = "2508.20229",
    archivePrefix = "arXiv",
    primaryClass = "astro-ph.HE",
    month = "8",
    year = "2025"
}

@article{Coronado-Blazquez:2021amj,
    author = "Coronado-Bl{\'a}zquez, Javier and S{\'a}nchez-Conde, Miguel A. and P{\'e}rez-Romero, Judit and Aguirre-Santaella, Alejandra",
    collaboration = "Fermi-LAT",
    title = "{Spatial extension of dark subhalos as seen by Fermi-LAT and implications for WIMP constraints}",
    eprint = "2204.00267",
    archivePrefix = "arXiv",
    primaryClass = "astro-ph.HE",
    doi = "10.1103/PhysRevD.105.083006",
    year = "2021"
}

@article{Fermi-LAT:2023zzt,
    author = "Smith, D. A. and others",
    collaboration = "Fermi-LAT",
    title = "{The Third Fermi Large Area Telescope Catalog of Gamma-Ray Pulsars}",
    eprint = "2307.11132",
    archivePrefix = "arXiv",
    primaryClass = "astro-ph.HE",
    doi = "10.3847/1538-4357/acee67",
    journal = "Astrophys. J.",
    volume = "958",
    number = "2",
    pages = "191",
    year = "2023"
}

@article{chernoff,
 ISSN = {00034851},
 URL = {http://www.jstor.org/stable/2236839},
 author = {Herman Chernoff},
 journal = {The Annals of Mathematical Statistics},
 number = {3},
 pages = {573--578},
 publisher = {Institute of Mathematical Statistics},
 title = {On the Distribution of the Likelihood Ratio},
 urldate = {2024-09-30},
 volume = {25},
 year = {1954}
}

@article{Bertone_2005,
   title={Particle dark matter: evidence, candidates and constraints},
   volume={405},
   ISSN={0370-1573},
   url={http://dx.doi.org/10.1016/j.physrep.2004.08.031},
   DOI={10.1016/j.physrep.2004.08.031},
   number={5–6},
   journal={Physics Reports},
   publisher={Elsevier BV},
   author={Bertone, Gianfranco and Hooper, Dan and Silk, Joseph},
   year={2005},
   month=jan, pages={279–390} }

@article{Bertone_2018,
   title={History of dark matter},
   volume={90},
   ISSN={1539-0756},
   url={http://dx.doi.org/10.1103/RevModPhys.90.045002},
   DOI={10.1103/revmodphys.90.045002},
   number={4},
   journal={Reviews of Modern Physics},
   publisher={American Physical Society (APS)},
   author={Bertone, Gianfranco and Hooper, Dan},
   year={2018},
   month=oct }

@article{Ajello_2016,
   title={FERMI-LAT OBSERVATIONS OF HIGH-ENERGY GAMMA-RAY EMISSION TOWARD THE GALACTIC CENTER},
   volume={819},
   ISSN={1538-4357},
   url={http://dx.doi.org/10.3847/0004-637X/819/1/44},
   DOI={10.3847/0004-637x/819/1/44},
   number={1},
   journal={The Astrophysical Journal},
   publisher={American Astronomical Society},
   author={Ajello, M. and others},
   year={2016},
   month=feb, pages={44} }

@article{Ackermann_2017,
   title={The Fermi Galactic Center GeV Excess and Implications for Dark Matter},
   volume={840},
   ISSN={1538-4357},
   url={http://dx.doi.org/10.3847/1538-4357/aa6cab},
   DOI={10.3847/1538-4357/aa6cab},
   number={1},
   journal={The Astrophysical Journal},
   publisher={American Astronomical Society},
   author={Ackermann, M. and others},
   year={2017},
   month=may, pages={43} }

@article{Di_Mauro_2021,
   title={Characteristics of the Galactic Center excess measured with 11 years of Fermi-LAT data},
   volume={103},
   ISSN={2470-0029},
   url={http://dx.doi.org/10.1103/PhysRevD.103.063029},
   DOI={10.1103/physrevd.103.063029},
   number={6},
   journal={Physical Review D},
   publisher={American Physical Society (APS)},
   author={Di Mauro, Mattia},
   year={2021},
   month=mar }

@article{Ackermann_2015,
   title={SEARCH FOR EXTENDED GAMMA-RAY EMISSION FROM THE VIRGO GALAXY CLUSTER WITH FERMI-LAT},
   volume={812},
   ISSN={1538-4357},
   url={http://dx.doi.org/10.1088/0004-637X/812/2/159},
   DOI={10.1088/0004-637x/812/2/159},
   number={2},
   journal={The Astrophysical Journal},
   publisher={American Astronomical Society},
   author={Ackermann, M. and others},
   year={2015},
   month=oct, pages={159} }

@article{Di_Mauro_2023,
   title={Constraining the dark matter contribution of 
gamma rays in clusters of galaxies using 
Fermi-LAT data},
   volume={107},
   ISSN={2470-0029},
   url={http://dx.doi.org/10.1103/PhysRevD.107.083030},
   DOI={10.1103/physrevd.107.083030},
   number={8},
   journal={Physical Review D},
   publisher={American Physical Society (APS)},
   author={Di Mauro, Mattia and Pérez-Romero, Judit and Sánchez-Conde, Miguel A. and Fornengo, Nicolao},
   year={2023},
   month=apr }

@misc{armand2021combineddarkmattersearches,
      title={Combined dark matter searches towards dwarf spheroidal galaxies with Fermi-LAT, HAWC, H.E.S.S., MAGIC, and VERITAS}, 
      author={Celine Armand and Eric Charles and Mattia di Mauro and Chiara Giuri and J. Patrick Harding and Daniel Kerszberg and Tjark Miener and Emmanuel Moulin and Louise Oakes and Vincent Poireau and Elisa Pueschel and Javier Rico and Lucia Rinchiuso and Daniel Salazar-Gallegos and Kirsten Tollefson and Benjamin Zitzer},
      year={2021},
      eprint={2108.13646},
      archivePrefix={arXiv},
      primaryClass={hep-ex},
      url={https://arxiv.org/abs/2108.13646}, 
}

@misc{mcdaniel2023legacyanalysisdarkmatter,
      title={Legacy Analysis of Dark Matter Annihilation from the Milky Way Dwarf Spheroidal Galaxies with 14 Years of Fermi-LAT Data}, 
      author={Alex McDaniel and Marco Ajello and Christopher M. Karwin and Mattia Di Mauro and Alex Drlica-Wagner and Miguel A. Sanchez-Conde},
      year={2023},
      eprint={2311.04982},
      archivePrefix={arXiv},
      primaryClass={astro-ph.HE},
      url={https://arxiv.org/abs/2311.04982}, 
}

@article{Ackermann_2012,
   title={SEARCH FOR DARK MATTER SATELLITES USINGFERMI-LAT},
   volume={747},
   ISSN={1538-4357},
   url={http://dx.doi.org/10.1088/0004-637X/747/2/121},
   DOI={10.1088/0004-637x/747/2/121},
   number={2},
   journal={The Astrophysical Journal},
   publisher={American Astronomical Society},
   author={Ackermann, M. and others},
   year={2012},
   month=feb, pages={121} }

@article{Ackermann_2018,
   title={The Search for Spatial Extension in High-latitude Sources Detected by the Fermi Large Area Telescope},
   volume={237},
   ISSN={1538-4365},
   url={http://dx.doi.org/10.3847/1538-4365/aacdf7},
   DOI={10.3847/1538-4365/aacdf7},
   number={2},
   journal={The Astrophysical Journal Supplement Series},
   publisher={American Astronomical Society},
   author={Ackermann, M. and others},
   year={2018},
   month=aug, pages={32} }

@misc{chou2017resolvingdarkmattersubhalos,
      title={Resolving Dark Matter Subhalos With Future Sub-GeV Gamma-Ray Telescopes}, 
      author={Ti-Lin Chou and Dimitrios Tanoglidis and Dan Hooper},
      year={2017},
      eprint={1709.08562},
      archivePrefix={arXiv},
      primaryClass={hep-ph},
      url={https://arxiv.org/abs/1709.08562}, 
}

@inbook{Balazs_2026,
   title={A primer on dark matter},
   ISBN={9780443214400},
   url={http://dx.doi.org/10.1016/B978-0-443-21439-4.00070-5},
   DOI={10.1016/b978-0-443-21439-4.00070-5},
   booktitle={Encyclopedia of Astrophysics},
   publisher={Elsevier},
   author={Balazs, Csaba and Bringmann, Torsten and Kahlhoefer, Felix and White, Martin},
   year={2026},
   pages={17–32} }

@article{Aguirre-Santaella:2023sww,
    author = "Aguirre-Santaella, Alejandra and S{\'a}nchez-Conde, Miguel A.",
    title = "{The viability of low-mass subhaloes as targets for gamma-ray dark matter searches}",
    eprint = "2309.02330",
    archivePrefix = "arXiv",
    primaryClass = "astro-ph.GA",
    doi = "10.1093/mnras/stae940",
    journal = "Mon. Not. Roy. Astron. Soc.",
    volume = "530",
    number = "3",
    pages = "2496--2511",
    year = "2024"
}

@article{Coronado_Bl_zquez_2022,
   title={Spatial extension of dark subhalos as seen by 
   Fermi-LAT and the implications for WIMP constraints},
   volume={105},
   ISSN={2470-0029},
   url={http://dx.doi.org/10.1103/PhysRevD.105.083006},
   DOI={10.1103/physrevd.105.083006},
   number={8},
   journal={Physical Review D},
   publisher={American Physical Society (APS)},
   author={Coronado-Blázquez, Javier and Sánchez-Conde, Miguel A. and Pérez-Romero, Judit and Aguirre-Santaella, Alejandra},
   year={2022},
   month=apr 
}

@article{Fermi-LAT:2022byn,
    author = "Abdollahi, Soheila and others",
    collaboration = "Fermi-LAT",
    title = "{Incremental Fermi Large Area Telescope Fourth Source Catalog}",
    eprint = "2201.11184",
    archivePrefix = "arXiv",
    primaryClass = "astro-ph.HE",
    doi = "10.3847/1538-4365/ac6751",
    journal = "Astrophys. J. Supp.",
    volume = "260",
    number = "2",
    pages = "53",
    year = "2022"
}

@article{Fermi-LAT:2017sxy,
    author = "Ajello, M. and others",
    collaboration = "Fermi-LAT",
    title = "{3FHL: The Third Catalog of Hard Fermi-LAT Sources}",
    eprint = "1702.00664",
    archivePrefix = "arXiv",
    primaryClass = "astro-ph.HE",
    doi = "10.3847/1538-4365/aa8221",
    journal = "Astrophys. J. Suppl.",
    volume = "232",
    number = "2",
    pages = "18",
    year = "2017"
}

@article{Fermi-LAT:2015uah,
    author = "Ackermann, M. and others",
    collaboration = "Fermi-LAT",
    title = "{2FHL: The Second Catalog of Hard Fermi-LAT Sources}",
    eprint = "1508.04449",
    archivePrefix = "arXiv",
    primaryClass = "astro-ph.HE",
    doi = "10.3847/0067-0049/222/1/5",
    journal = "Astrophys. J. Suppl.",
    volume = "222",
    number = "1",
    pages = "5",
    year = "2016"
}

@article{Fermi-LAT:2015bhf,
    author = "Acero, F. and others",
    collaboration = "Fermi-LAT",
    title = "{Fermi Large Area Telescope Third Source Catalog}",
    eprint = "1501.02003",
    archivePrefix = "arXiv",
    primaryClass = "astro-ph.HE",
    doi = "10.1088/0067-0049/218/2/23",
    journal = "Astrophys. J. Suppl.",
    volume = "218",
    number = "2",
    pages = "23",
    year = "2015"
}

@ARTICLE{2009ApJ...699.1171A,
       author = {{Abdo}, A.~A. and others},
        title = "{Pulsed Gamma Rays from the Millisecond Pulsar J0030+0451 with the Fermi Large Area Telescope}",
      journal = {apj},
         year = 2009,
        month = jul,
       volume = {699},
       number = {2},
        pages = {1171-1177},
          doi = {10.1088/0004-637X/699/2/1171},
archivePrefix = {arXiv},
       eprint = {0904.4377},
 primaryClass = {astro-ph.HE},
       adsurl = {https://ui.adsabs.harvard.edu/abs/2009ApJ...699.1171A}
}

@article{Pathania:2025iyj,
    author = "Pathania, A. and Singh, K. K. and Singh, S. K. and Tolamatti, A. and Singh, B. B. and Yadav, K. K.",
    title = "{Identification of gamma ray pulsar candidates in the Fermi-LAT 4FGL-DR4 unassociated sources using supervised machine learning}",
    eprint = "2510.08654",
    archivePrefix = "arXiv",
    primaryClass = "astro-ph.HE",
    doi = "10.1016/j.astropartphys.2025.103185",
    journal = "Astropart. Phys.",
    volume = "175",
    pages = "103185",
    year = "2026"
}

@article{Manchester:2017azn,
    author = "Manchester, R. N.",
    title = "{Millisecond Pulsars, their Evolution and Applications}",
    eprint = "1709.09434",
    archivePrefix = "arXiv",
    primaryClass = "astro-ph.HE",
    doi = "10.1007/s12036-017-9469-2",
    journal = "J. Astrophys. Astron.",
    volume = "38",
    pages = "42",
    year = "2017"
}

@article{crawford2005search,
  author       = {Crawford, Fronefield and McLaughlin, Maura and Johnston, Simon and Romani, Roger and Sorrelgreen, Ethan},
  title        = {A Search for Radio Emission from the Young 16-ms X-ray Pulsar PSR~J0537{-}6910},
  journal      = {Advances in Space Research},
  volume       = {35},
  number       = {1181},
  pages        = {1181--1184},
  year         = {2005},
  doi          = {10.1016/j.asr.2005.03.074},
  arxiv        = {astro-ph/0503516}
}

@ARTICLE{2016MNRAS.458L..84A,
       author = {{Algeri}, Sara and {Conrad}, Jan and {van Dyk}, David A.},
        title = "{A method for comparing non-nested models with application to astrophysical searches for new physics}",
      journal = {mnras},
         year = 2016,
        month = may,
       volume = {458},
       number = {1},
        pages = {L84-L88},
          doi = {10.1093/mnrasl/slw025},
archivePrefix = {arXiv},
       eprint = {1509.01010},
 primaryClass = {physics.data-an},
       adsurl = {https://ui.adsabs.harvard.edu/abs/2016MNRAS.458L..84A}
}

@article{DelaTorreLuque:2024ozf,
    author = "De la Torre Luque, Pedro and Winkler, Martin Wolfgang and Linden, Tim",
    title = "{Antiproton bounds on dark matter annihilation from a combined analysis using the DRAGON2 code}",
    eprint = "2401.10329",
    archivePrefix = "arXiv",
    primaryClass = "astro-ph.HE",
    doi = "10.1088/1475-7516/2024/05/104",
    journal = "JCAP",
    volume = "05",
    pages = "104",
    year = "2024"
}

@article{BurnhamAnderson2004,
  author  = {Burnham, Kenneth P. and Anderson, David R.},
  title   = {Multimodel Inference: Understanding AIC and BIC in Model Selection},
  journal = {Sociological Methods \& Research},
  year    = {2004},
  volume  = {33},
  number  = {2},
  pages   = {261--304},
  doi     = {10.1177/0049124104268644}
}

@misc{arina2023cosmixscosmicmessengerspectra,
      title={CosmiXs: Cosmic messenger spectra for indirect dark matter searches}, 
      author={Chiara Arina and Mattia Di Mauro and Nicolao Fornengo and Jan Heisig and Adil Jueid and Roberto Ruiz de Austri},
      year={2023},
      eprint={2312.01153},
      archivePrefix={arXiv},
      primaryClass={astro-ph.HE},
      url={https://arxiv.org/abs/2312.01153}, 
}

@inproceedings{Bruel:2018lac,
    author = "Bruel, P. and Burnett, T. H. and Digel, S. W. and Johannesson, G. and Omodei, N. and Wood, M.",
    collaboration = "Fermi-LAT",
    title = "{Fermi-LAT improved Pass{\textasciitilde}8 event selection}",
    booktitle = "{8th International Fermi Symposium}: {Celebrating 10 Year of Fermi}",
    eprint = "1810.11394",
    archivePrefix = "arXiv",
    primaryClass = "astro-ph.IM",
    month = "10",
    year = "2018"
}

\end{document}